\documentclass[fleqn,usenatbib,useAMS]{mnras}

\usepackage{graphicx}	% Including figure files
\usepackage{amsmath}	% Advanced maths commands
\usepackage{amssymb}	% Extra maths symbols
\usepackage{multicol}        % Multi-column entries in tables
\usepackage{bm}		% Bold maths symbols, including upright Greek
\usepackage{pdflscape}	% Landscape pages
\newcommand{\vvec}[1]{{\textbf{#1}}}

\newcommand{\gx}{\textsc{Gadget-X}}

\newcommand{\ahf}{\textsc{AHF}}

\newcommand{\Mbnd}{{\ifmmode{M_{\rm bnd}}\else{$M_{\rm bnd}$}\fi}}
\newcommand{\Mfof}{{\ifmmode{M_{\rm fof}}\else{$M_{\rm fof}$}\fi}}
\newcommand{\Mcrit}{{\ifmmode{M_{\rm 200c}}\else{$M_{\rm 200c}$}\fi}}
\newcommand{\Rcrit}{{\ifmmode{R_{\rm 200c}}\else{$R_{\rm 200c}$}\fi}}
\newcommand{\Rhost}{{\ifmmode{R_{\rm host}}\else{$R_{\rm host}$}\fi}}
\newcommand{\Mmean}{{\ifmmode{M_{\rm 200m}}\else{$M_{\rm 200m}$}\fi}}
\newcommand{\MBN}{{\ifmmode{M_{\rm BN98}}\else{$M_{\rm BN98}$}\fi}}

\newcommand{\hGpc}{{\ifmmode{h^{-1}{\rm Gpc}}\else{$h^{-1}$Gpc}\fi}}
\newcommand{\hMpc}{{\ifmmode{h^{-1}{\rm Mpc}}\else{$h^{-1}$Mpc}\fi}}
\newcommand{\hkpc}{{\ifmmode{h^{-1}{\rm kpc}}\else{$h^{-1}$kpc}\fi}}
\newcommand{\hMsun}{{\ifmmode{h^{-1}{\rm {M_{\odot}}}}\else{$h^{-1}{\rm{M_{\odot}}}$}\fi}}
\newcommand{\Mstar}{{\ifmmode{M_{*}}\else{$M_{*}$}\fi}}
\newcommand{\Mhalo}{{\ifmmode{M_{\rm Halo}}\else{$M_{\rm Halo}$}\fi}}
\newcommand{\Ngal}{{\ifmmode{N_{\rm gal}}\else{$N_{\rm gal}$}\fi}}
\newcommand{\Norph}{{\ifmmode{N_{\rm orphan}}\else{$N_{\rm orphan}$}\fi}}
\newcommand{\Nxorph}{{\ifmmode{N_{\rm non-orphan}}\else{$N_{\rm non-orphan}$}\fi}}
\newcommand{\Zsolar}{{\ifmmode{Z_{\odot}}\else{$Z_{\odot}$}\fi}}
\newcommand{\Msun}{{\ifmmode{{\rm {M_{\odot}}}}\else{${\rm{M_{\odot}}}$}\fi}}
\newcommand{\ltsima}{$\; \buildrel < \over \sim \;$}
\newcommand{\gtsima}{$\; \buildrel > \over \sim \;$}
\newcommand{\lsim}{\lower.5ex\hbox{\ltsima}}
\newcommand{\gsim}{\lower.5ex\hbox{\gtsima}}

\newcommand{\Tab}[1]{Table~\ref{#1}}
\newcommand{\Sec}[1]{Section~\ref{#1}}
\newcommand{\App}[1]{Appendix~\ref{#1}}
\newcommand{\Eq}[1]{Eq.~(\ref{#1})}
\newcommand{\Fig}[1]{Fig.~\ref{#1}}
\newcommand{\beq}{\begin{equation}}
\newcommand{\eeq}{\end{equation}}

\usepackage[T1]{fontenc}
\usepackage{ae,aecompl}

\usepackage{newtxtext,newtxmath}
\title[Radial Alignment in Galaxy Clusters]{The Three Hundred project: shapes and radial alignment of satellite, infalling, and backsplash galaxies}

\author[Knebe et al.]
{Alexander Knebe,$^{1,2,3}$\thanks{Contact e-mail: \href{mailto:alexander.knebe@uam.es}{alexander.knebe@uam.es}}
Mat\'{i}as G\'{a}mez-Mar\'{i}n,$^{1}$
Frazer R. Pearce,$^4$
Weiguang Cui,$^5$
\newauthor
Kai Hoffmann,$^6$
Marco De Petris,$^{7,8}$
Chris Power,$^3$
Roan Haggar,$^4$
Robert Mostoghiu$^1$
\\
$^{1}$Departamento de F\'isica Te\'{o}rica, M\'{o}dulo 15, Facultad de Ciencias, Universidad Aut\'{o}noma de Madrid, 28049 Madrid, Spain\\
$^{2}$Centro de Investigaci\'{o}n Avanzada en F\'isica Fundamental (CIAFF), Facultad de Ciencias, Universidad Aut\'{o}noma de Madrid, 28049 Madrid, Spain\\
$^{3}$International Centre for Radio Astronomy Research, University of Western Australia, 35 Stirling Highway, Crawley, Western Australia 6009, Australia\\
$^{4}$School of Physics \& Astronomy, University of Nottingham, Nottingham NG7 2RD, UK\\
$^{5}$Institute for Astronomy, University of Edinburgh, Royal Observatory, Edinburgh EH9 3HJ, United Kingdom\\
$^{6}$Institute for Computational Science, University of Zurich, Winterthurerstr. 190, 8057 Z\"urich, Switzerland\\
$^{7}$Department of Physics, Sapienza Universit\`{a} di Roma, p.le Aldo Moro 5, I-00185 Rome, Italy\\
$^{8}$INFN - Sezione di Roma, P.le A. Moro 2, I-00185 Roma, Italy\\
}
\date{Last updated 2015 May 22; in original form 2013 September 5}

\pubyear{2020}

\begin{document}
\label{firstpage}
\pagerange{\pageref{firstpage}--\pageref{lastpage}}
\maketitle

\begin{abstract}
Using 324 numerically modelled galaxy clusters we investigate the radial and galaxy-halo alignment of dark matter subhaloes and satellite galaxies  orbiting within and around them. We find that radial alignment depends on distance to the centre of the galaxy cluster but appears independent of the dynamical state of the central host cluster. Furthermore, we cannot find a relation between radial alignment of the halo or galaxy shape with its own mass. We report that backsplash galaxies, i.e. objects that have already passed through the cluster radius but are now located in the outskirts, show a stronger radial alignment than infalling objects. We further find that there exists a population of well radially aligned objects passing very close to the central cluster's centre which were found to be on highly radial orbit.
\end{abstract}


%%%%%%%%%%%%%%%%%%%%%%%%%%%%%%%%%%%%%%%%%%%%%%%%%%

%%%%%%%%%%%%%%%%% BODY OF PAPER %%%%%%%%%%%%%%%%%%

%%%%%%%%%%%%%%%%%%%%%%%%%%%%%%%%%%%%%%%%%%%%%%%%%%
\section{Introduction}
%%%%%%%%%%%%%%%%%%%%%%%%%%%%%%%%%%%%%%%%%%%%%%%%%%
Weak gravitational lensing caused by the large-scale structure of the Universe (i.e. the `cosmic shear') induces correlations in the observed shapes of galaxies. For surveys covering a large enough fraction of the sky, such a signal can then be used to obtain valuable information about structure formation in the Universe and eventually cosmological parameters. However, this requires a profound and detailed understanding of natural galaxy alignments, i.e. alignments not apparently induced by lensing but caused by other mechanisms such as, for instance, cosmic structure formation itself. This is in particular important at small scales, where the signal-to-noise ratio of weak lensing statistics is highest. Therefore, the last decade has seen a lot of work aiming at understanding the origin of galaxy alignments such as correlations between the galaxy and halo shape, the orientation of galaxy shapes with respects to their local cosmic web, and the relation between galaxy position and its shape. For an exhaustive introduction to the field and an overview of all these phenomena we refer the reader to the two elaborate review articles by \citet{Joachimi2015} and \citet{Kiessling2015}.

In this work we focus on the alignment of galaxies with respect to the host halo they orbit. This topic has already been addressed from three different perspectives: observations, analytical models, and numerical simulations. Here we are going to concentrate our efforts on the latter. We further focus on one particular type of alignment, i.e. the `radial alignment' (sometimes also referred to as `shape alignment'\footnote{In some observational studies this is also referred to as `satellite alignment'.}). This alignment -- as depicted in \Fig{fig:angles} below -- measures the correlation between the major axis of the elliptical shape of a galaxy (or its dark matter halo) to its position with respects to the centre of the nearest larger object (which most commonly is the galaxy cluster in which the galaxy orbits). However, we will also briefly touch upon the `galaxy-halo alignment' that evaluates the orientation between the shape of a galaxy and its own dark matter halo.

Radial alignment has been studied in the 2000's by means of dark matter only simulations and an appreciable signal has been found for halo shapes \citep[][]{Kuhlen07,Pereira08,Faltenbacher08,Knebe08,Knebe08b}. But to date only a few works exist that are based upon cosmological simulations that also include all the relevant baryonic effects. To our knowledge the only contributions including baryon physics are the studies of \citet{Knebe10a}, \citet{Tenneti2015b}, \citet{Velliscig15b}, and \citet{Chisari2017}. However, \citet{Barber2015} used a semi-analytical modelling approach -- as opposed to full physics hydrodynamical simulations -- to investigate the alignment of dwarf spheroidal galaxies within the (dark matter only) `Aquarius' simulation suite. 

While radial alignment has been found in cosmological simulations whether sub-grid physics have been included or not, the situation is less clear for observations. Parallel to the numerical modelling in the 2000's, the utilisation of early SDSS data confirmed that the major axis of galaxies preferentially points towards the centre of the cluster they reside in \citep[e.g.][]{Pereira05,Agustsson06}. However, these findings have recently been challenged as the majority of newer observations tend to either indicate no such alignment \citep[e.g.][]{Hung2012,Schneider2013,Chisari2014,Sifon2015} or only alignment for the most luminous galaxies \citep{Singh2015}. However, very recent works not only reports that such a signal exists \citep[][]{Wang2019,Pajowska2019}, but also detected a dependence on distance to the central object \citep[][]{Huang2018,Georgiou2019}.

The radial alignment of satellites and the alignment between the dark matter halo and galaxies are important ingredients for toy models of galaxy shapes as found in large volume cosmological dark matter only simulations \citep[][]{Joachimi2013a,Joachimi2013b}, which in turn are being used to quantify the aforementioned contamination of weak lensing surveys. We add that the halo model for intrinsic alignment in weak lensing analyses assumes that satellites point towards central galaxies \citep{Troxel2015}, and it fits observations rather well \citep[see, for instance,][]{Singh2015}. The analysis presented here will hence help to understand how to improve these models using more realistic assumptions on the mass and scale dependence of the galaxy-halo alignment.

To investigate the presence (or not) of radial alignment in cluster galaxiy members, we employ `The Three Hundred' data set that consists of regions of diameter $30\hMpc$ centred on the 324 most massive objects found within a cosmological dark matter only simulation of side length $1000\hMpc$. Those regions have been re-simulated with \gx, i.e. full physics hydrodynamical code for cosmological simulations based upon a modern SPH solver. For more details about both the code and the general data set, we refer the reader to the paper by \citet{Cui18}. When analysing our data we then put a special focus on the differences found for radial alignments of subhaloes, infalling, and backsplash objects. We are particularly interested in contrasting the latter two populations, i.e. those objects that are approaching the host galaxy cluster for the first time and those that already passed through its radius. 

The paper is structured as follows. In \Sec{sec:data} we introduce our data set and the way we calculate halo and galaxy shapes as well as how to quantify alignment. In \Sec{sec:shapes_and_IA} we then present the distribution of shapes of our objects; we also briefly touch upon `galaxy-halo' that measures the orientation of the shape of the stellar component with respects to that of the host halo it resides in. Our main results are then shown in \Sec{sec:radialalignment} where we study the radial alignment and its relation to several other factors (e.g. distance to central galaxy clusters, etc.). We conclude in \Sec{sec:conclusions}.

%%%%%%%%%%%%%%%%%%%%%%%%%%%%%%%%%%%%%%%%%%%%%%%%%%
\section{The Data} \label{sec:data}
%%%%%%%%%%%%%%%%%%%%%%%%%%%%%%%%%%%%%%%%%%%%%%%%%%

\subsection{`The Three Hundred' Galaxy Clusters} \label{sec:simulations}
%%%%%%%%%%%%%%%%%%%%%%%%%%%%%%%%%%%%%%%%%%%%%%%%%%
Our data constitutes of `The Three Hundred' theoretically modelled galaxy clusters.\footnote{\url{http://www.the300-project.org}} The 324 objects -- simulated using a so-called pseudo-zoom technique\footnote{The 324 most massive objects found in the dark-matter only MDPL2 simulation (cf. \url{http://www.cosmosim.org}) have been selected and a 15\hMpc\ region about them populated with gas particles; the simulation was then re-run, but particles outside that region sequentially degraded in mass resolution. The re-simulation then modelled all relevant baryonic physics for those gas particles and hence we end up with a full halo and galaxy catalogue for each of the 324 central galaxy clusters within a 15\hMpc\ sphere.} -- form a mass-complete set covering the range $M_{\rm 200} \in [6.4, 26.5] \times 10^{14}\hMsun$ (at redshift $z=0$) and have been presented in \citet{Cui18}. One of the strengths of our data is that the size of our sample that allows statistically significant subsamples to be constructed. As all the details of the models can be found in either the introductory paper by \citet{Cui18} or any of the other already published papers based upon this data \citep[i.e.][]{Mostoghiu18,Wang18,Arthur19,Ansarifard2020,Haggar2020} we are only going to briefly highlight here the prime aspects of the \gx\ code that has been used to model our data.

\paragraph*{\gx} This is an advanced version of \textsc{Gadget3} incorporating an improved SPH scheme with artificial thermal diffusion, time-dependent artificial viscosity, high-order Wendland C4 interpolating kernel, and wake-up scheme \citep{Beck2016}. Star formation follows the classical \citet{Springel03} prescription and is implemented in a stochastic way which leads to varying star particle masses of order $m_{*}\sim 4\times 10^{7}\hMsun$. Stellar evolution and metal enrichment is originally described in \citet{Tornatore2007} with further updates in \citet{Murante2010} and \citet{Rasia2015}. It further implements the black hole growth and AGN feedback of \citet{Steinborn2015}.
%%%%%%%%%%%%%%%%%%%%%%%%%%%%%%%%%%%%%%%%%%%%%%%%%%

\subsection{Shape Calculation} \label{sec:shapecalculation}
%%%%%%%%%%%%%%%%%%%%%%%%%%%%%%%%%%%%%%%%%%%%%%%%%%
All our objects have been identified with the open-source halo finder \ahf\footnote{\ahf\ can be freely downloaded from \url{http://popia.ft.uam.es/AHF}.} \citep{Knollmann09,Gill04a}. \ahf\ locates density peaks in an adaptively smoothed density field of the simulation collecting particles gravitationally bound to it. In that process it not only considers dark matter but also star and gas particles where for the latter the thermal energy is taken into account during the unbinding procedure. All objects (which might include stellar- and/or gas-only objects) with at least 20 particles are kept and a suite of integral properties calculated. Please note that while the radius of field haloes is calculated via ${M_{\rm 200c}(<\Rcrit)} = {\displaystyle 200 \rho_{\rm crit} \Rcrit^3} 4\pi/3$, subhaloes could have their radius truncated earlier due to the embedding within the background density of their host halo. For \ahf\ subhaloes the radius (and hence the particles considered for the shape calculation) is eventually determined as the distance to the farthest gravitationally bound particle \citep[see][]{Knollmann09}.

For the calculation of the shapes (and the respective orientations of them) we are utilizing the so-called reduced moment of inertia tensor

\begin{equation} \label{eq:RMIT}
 I_{i,j}=\sum_nm_n x_{i,n}x_{j,n}/r_n^2 \ ,
\end{equation}

\noindent
where $x_{i,n}$ and $x_{j,n}$ are the $i$th and $j$th component of the $n$th particle coordinate and $r_n$ the particle's distance to the centre of the object; note that -- due to the $1/r^2$ weighting -- the reduced version puts more emphasis onto the central region which is especially important when aiming at determining the shape of the galaxy in the central part of the objects. Our determination of the eigenvalues $a>b>c$ used for the shape determination is based upon a diagonalisation of $I_{i,j}$ which will also equip us with the corresponding eigenvectors $\vvec{e}_{a}, \vvec{e}_{b}, \vvec{e}_{c}$. We define sphericity as the ratio between the smallest and largest eigenvalue

\begin{equation}
 s=\frac{c}{a},
\end{equation}

\noindent
and primarily consider the major axis $\vvec{e}_{a}$ for the study of alignment. Note that we will distinguish two reduced moment of inertia tensors, one that is based upon \textit{all} particles within the haloes' radius (incl. dark matter, gas, and star particles) and one that uses \textit{only star particles} inside the halo. The corresponding sphericities and axes will be superscripted $h$ (for `halo', i.e. all particles) and $*$ (for `stellar', i.e. only star particles), respectively.

The determination of the halo centres might impact the measurement of the inertia tensor, especially for irregular-shaped halos. We therefore like to mention that the centres returned by our halo finder are `peaks in the density field'. And as the density contours are derived via an adaptive-mesh refinement technique \citep[see][]{Knollmann09} they are arbitrarily shaped, perfectly following the actual density field. We therefore do not expect our density-based centre determination to be affected by irregularly shaped objects. The performance of the centre determination for spherical NFW haloes can, for instance, be seen in \citet[Fig.2 of][]{Knebe11}. Further note, there is only one centre for both the halo and `galaxy' shape, i.e. the one based upon the whole matter density field (which includes the stars and gas).

We like to close with two remarks of caution. First, due to the spherical overdensity nature of our halo finder, shapes will be biased towards larger sphericities. An in-depth study of this situation has been presented in \citet[][]{Bailin05} where they find that using spherical shells, rather than ellipsoids defined by isodensity contours, does not affect the orientation of the principal axes; it does, however, bias the derived axis ratios toward spherical. They further provide a formula to post-correct for this bias (E4.(4) in their paper, based upon their Fig.1). In order to leave our own results as `lower limits' we chose to not correct for the bias (but please see \Fig{fig:ShapeMassCorrected} in \App{app:supplements}). Second, using all particles when calculating the overall halo shape includes again the stellar particles and hence might give a biased estimate. But as has been shown in previous works, the calculation will be dominated by the dark matter -- if applying an upper limit for the stellar-to-halo mass (SMHM) ratio -- and hence this does not affect the results \cite[cf. Fig.~1 in][]{Knebe10a}. We finally remark that all our analyses will be performed in 3D to profit from the full information contained in the simulations.

%\begin{savenotes}
\begin{table*}
  \caption{Number of subhaloes ($N_{\rm s}$, first column), backsplash ($N_{\rm b}$, second column), infalling ($N_{\rm i}$, third column) and all objects ($N_{\rm all}$, fourth column) as found by the halo finder within $3\times \Rcrit$, and successively applying our selection criteria, i.e. stellar mass threshold ($M_{*}\geq 10^{11}\hMsun$), restriction of stellar-to-halo mass (SMHM) ratio ($M_*/M_{\rm halo}<0.2$), removal of sub-substructure, `exclusiveness' ($M_{*}^{\rm ex}/M_{*}^{\rm in}>0.85$), and prolateness $b/a$. The last column features the range of halo masses covered by the selected objects.}
	\label{tab:number_of_objects}\vspace{-0.2cm}
	\setlength{\tabcolsep}{1.0pt}
		\begin{center}
			\begin{tabular}{lccccc}
			\hline
			selection criterion             & $N_{\rm s}$   & $N_{\rm b}$   & $N_{\rm i}$   & $N_{\rm all}$         & range of halo masses $M_{\rm h}$ [$h^{-1}$M$_{\odot}$] \\
			\hline
			none                            & 144651        & 66050         & 148286        & 358987    & 7.4$\cdot$10$^{8}$ - 1.1$\cdot$10$^{15}$  \\
			mass cut                        & 3586          & 764           & 1840          & 6190        & 1.1$\cdot$10$^{11}$ - 1.1$\cdot$10$^{15}$  \\
			+ SMHM ratio restriction        & 3073          & 752           & 1801          & 5626        & 5.0$\cdot$10$^{11}$ - 1.1$\cdot$10$^{15}$  \\
			+ sub-substructure removal      & 2645          & 741           & 1551          & 4937                  & 5.0$\cdot$10$^{11}$ - 1.1$\cdot$10$^{15}$  \\
			+ `exclusiveness' restriction   & 2371          & 679           & 781           & 3831                  & 5.0$\cdot$10$^{11}$ - 5.3$\cdot$10$^{14}$  \\
			+ $b/a<0.9$ selection           &    1767   &        373       &         306      &      2446         &  5.0$\cdot$10$^{11}$ - 5.3$\cdot$10$^{14}$  \\
			\hline
			\end{tabular}
		\end{center}
	\end{table*}
%\end{savenotes}

\subsection{Object Selection} \label{sec:objectselection}
%%%%%%%%%%%%%%%%%%%%%%%%%%%%%%%%%%%%%%%%%%%%%%%%%%
We consider all objects out to $3\times\Rcrit$ where \Rcrit\ is the radius of the central galaxy cluster and defined via
\begin{equation}
 M(<\Rcrit)=200 \ \rho_{\rm crit} \ 4\pi R^3_{\rm 200c}/3 .
\end{equation}

\noindent
Note that for all objects considered here the reference frame is given by the central galaxy cluster. In view of various technical limitations related to, for instance the need to include a sufficiently large number of particles when calculating shapes, we are eventually not using all objects found by the halo finder. We are restricting our sample by applying several selection criteria detailed in the following.

\paragraph*{Mass limit} \ahf\ provides objects with as few as 20 particles. However, this number is too low to infer reliable shape measurements. While it has been advocated that 200-300 particles are sufficient \citep[e.g.][]{Pereira08,Knebe10a,Velliscig15b,Chisari2017}, the convergence studies of \citet{Tenneti14} and \citet[][Figs. A1 and A2]{Hoffmann2014} argue for a more conservative value of 1000 particles. We decided to only use objects with a stellar mass of at least $M_{*}\geq 10^{11}\hMsun$. Note, the stellar particles in \gx\ have varying masses due to the nature of the implementation of sub-grid physics, but our mass cut ensures we always have $N_{*}\geq 1000$ and hence $N_{\rm h}>1000$, too.

\paragraph*{Stellar-to-halo mass ratio limit} As mentioned in \Sec{sec:shapecalculation} our halo shape calculations are based upon \textit{all} particles within the object and hence also include the star particles which are the ones used for the galaxy shape determination, too. But if the ratio between stellar mass $M_{*}$ and halo mass $M_{\rm halo}=M_{\rm dm}+M_{*}+M_g$ approaches unity\footnote{Remember that \ahf\ finds all gravitationally bound particle aggregations and some of those are in fact made up of objects primarily consisting in stellar particles \citep[see][]{Mostoghiu20}.} this then entails that both halo and galaxy shape will be identical. To avoid this we are restricting the stellar-to-halo mass ratio to be $M_*/M_{\rm halo}<0.2$ (cf. Fig.6 in \citet{Cui18}).

\paragraph*{Sub-subhaloes} Our cluster simulations have high enough mass resolution to actually resolve sub-subhaloes, i.e. haloes orbiting \textit{within} the objects for which we determine alignment. But as our main focus lies with the alignment of subhaloes with respects to the central galaxy cluster we are removing those sub-subhaloes from our study. They would need to be studied in the rest-frame of their respective subhalo frame.

\paragraph*{Inclusive vs. exclusive particles} The particle content of all our \ahf\ haloes is `inclusive', meaning that substructure is considered part of its host halo. This implies that particles in the substructure also contribute to the shape of the host halo.\footnote{Note that what is considered `host' halo here might as well be a sub-halo of the central galaxy cluster.} And even though the $1/r^2$ weighting in the reduced moment of inertia tensor corrects for that, it might affect massive objects in the outer regions (i.e. beyond \Rcrit) of the central galaxy cluster. Their sphericities $s^{h/*}$ as determined using the `inclusive' particles are smaller (i.e. less spherical) than the sphericity as measured by, for instance, only the innermost 20 per cent of particles.\footnote{This has been verified for a small yet representative sample of affected objects by extracting their (inclusive) particles from the original simulation and calculating the radial profile of sphericity $s^{h/*}(r)$.} In order to remove affected objects we employ the following strategy: for the stellar component, \ahf\ not only returns the inclusive $M_{*}^{\rm in}$ but also the exclusive mass $M_{*}^{\rm ex}$ for all objects. Therefore, the ratio $M_{*}^{\rm ex}/M_{*}^{\rm in}$ provides us with a means to quantify the corruption of the shape calculation. Restricting this ratio to be larger than 0.85 leaves us with clean objects -- although reducing it to even 0.5 does not affect any of the radial alignment plots presented below. We further remark that the stellar mass ratio criterion is more restrictive than the substructure fraction limitation.\\

\paragraph*{Prolateness} For oblate objects (characterized by $a\sim b > c$) there exist a degeneracy between the two major axes. Therefore, using angles that involve the major axis $a$ might result in unclear signals unless a `prolateness criterion' is applied. We therefore restrict the analysis to objects with axis ratios $b/a<0.9$ as advocated in \citet{Pereira08} and \citet{Knebe08}.\\

\subsection{Object Populations}
%%%%%%%%%%%%%%%%%%%%%%%%%%%%%%%%%%%%%%%%%%%%%%%%%%%%
For the remainder of the work we will distinguish three distinct populations of objects defined as follows:

\begin{itemize}
    \item[a)] \textit{subhaloes}: objects located at redshift $z=0$ inside $\Rcrit$,
    \item[b)] \textit{infalling haloes}: these are objects that have not yet entered \Rcrit\ and are on their first passage towards the host cluster,
    \item[c)] \textit{backsplash galaxies}: these objects have already passed through \Rcrit, but are now outside of $\Rcrit$; please refer to \citet{Haggar2020} for an exhaustive discussion of their properties.
\end{itemize}

\noindent
This distinction allows us to check for and quantify the contribution of objects that already were under the (tidal) influence of the host halo to the radial alignment signal in the outskirts of the cluster.\\

In order to allow for this division we are tracking the orbits of all our objects backwards in time as described in detail in a companion paper \citep{Haggar2020}. In doing so we also record the cases where a halo actually crosses \Rcrit\ multiple times. Note, an infalling object has not yet crossed \Rcrit, a subhalo normally crossed once, and a backsplash galaxy twice. But we also find few instances of more than two crossings: 12 per cent of all the objects considered traversed \Rcrit\ more than twice, but 85 per cent of those eventually ended up as subhaloes at redshift $z=0$ (the other 15 per cent are obviously backsplash haloes). We do not believe that it adds to the results to study them separately and hence refrain from doing so.\\

 In all that follows we are stacking our selected objects for all 324 central galaxy clusters. We summarize the reduction of the number of objects due to successive application of the selection criteria in \Tab{tab:number_of_objects}. Besides of listing the number of subhaloes (i.e. objects within \Rcrit, first column) we also give the number of backsplash (second column) and infalling (third column) objects out to $3\times\Rcrit$; the fourth column is simply the sum of the three previous columns, and in the last column we provide the mass range of all objects. We like to remark that practically all backsplash haloes are situated in $[\Rcrit,2 \Rcrit]$ with approximately equal numbers of infalling objects in $[\Rcrit,2 \Rcrit]$ and $[2 \Rcrit,3 \Rcrit]$ \citep[cf.][]{Haggar2020}. We notice that the most stringent selection is given by the stellar mass cut; all other cuts just marginally lower the number of objects. However, the exception to that rule is found for the infalling haloes when applying the `exclusiveness' restriction. This criterion basically checks for sub-structure in the objects of interest, and the sharp drop when applying it to the infalling population indicates that nearly half of that population contains sub-structure and falls towards the main central galaxy cluster in groups, respectively. We will investigate such `group infall' in more detail in a companion paper (Haggar et al., in prep.).\\
 
 For clarity we like to briefly summarize the nature of the objects considered in the following analysis. Our reference frame for distances and velocities is the halo of the central galaxy cluster of which there are 324 in our \textsc{The Three Hundred} data set. Applying all the aforementioned selection criteria to all objects found within $3\times\Rcrit$, we now distinguish between objects inside (subhaloes) and outside (either infalling or backsplash) the central cluster's radius \Rcrit. Each of these objects contains dark matter, stars, and gas. We then study the alignment of both the shape of the total and stellar matter content with respects to position and velocity in the central cluster's rest frame (referred to as `radial alignment'). We also check the relation between the shape of the total and stellar matter itself (referred to as `galaxy-halo' alignment).\\

 \begin{figure}
 \begin{center}
   \includegraphics[width=7cm]{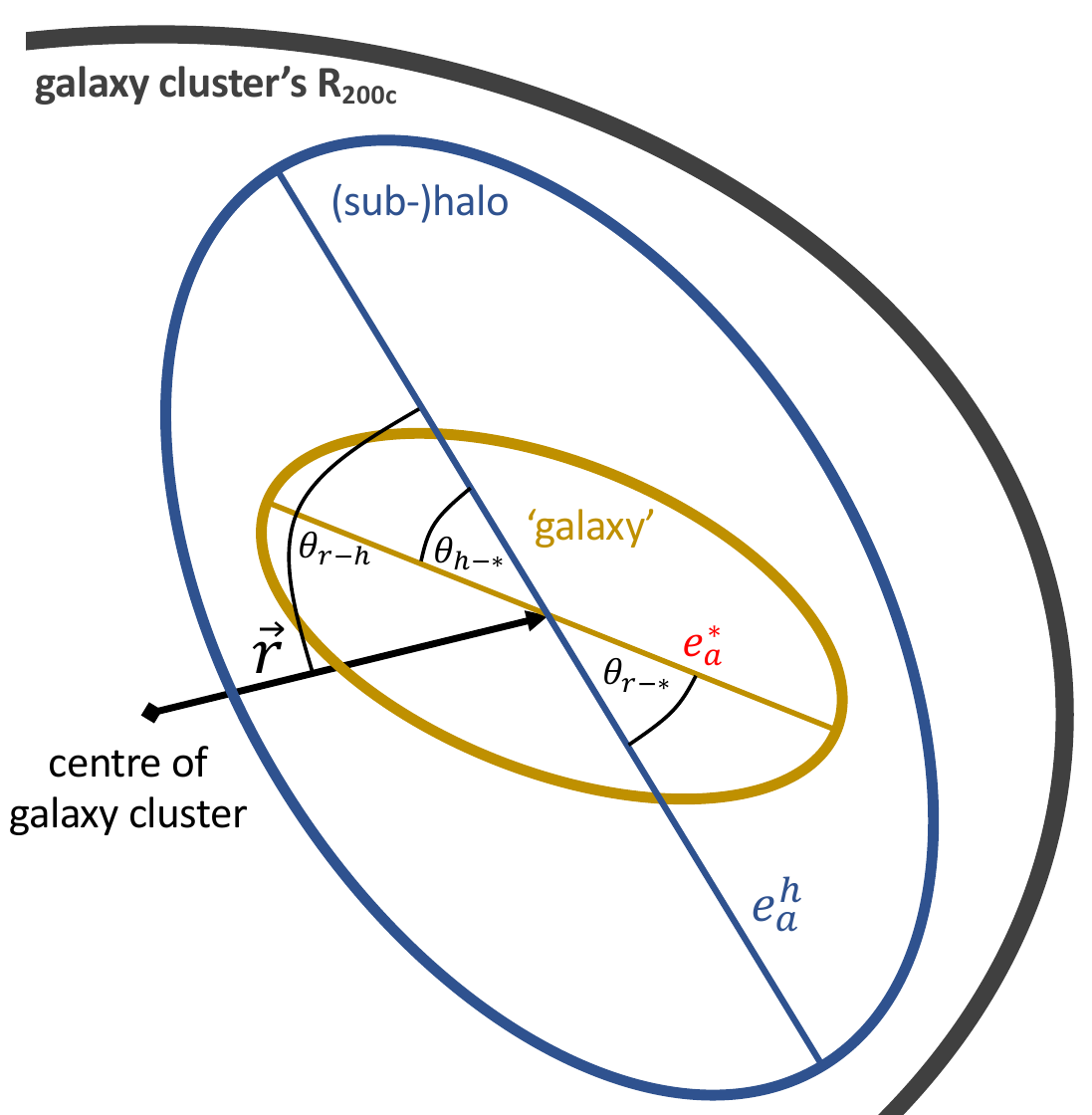}
 \end{center}
   \caption{Illustration of the eigenvectors and angles used for the study of alignments.}
 \label{fig:angles}
 \end{figure}

\subsection{Alignments}\label{sec:alignments}
%%%%%%%%%%%%%%%%%%%%%%%%%%%%%%%%%%%%%%%%%%%%%%%%%%
Previous theoretical studies have reported that the distribution of the different structures found within galaxy clusters is not random. In fact, one can infer that there is a tendency of sub-structures to have their intrinsic shapes oriented towards preferential directions, which will depend on the environment that embeds them \citep[e.g.][]{Kuhlen07,Pereira08,Faltenbacher08,Knebe08,Knebe10a,Tenneti2015b,Velliscig15b}. To measure such alignments it is common practice to approximate the ellipsoidal shape of the objects by using the eigenvectors of the reduced moment of inertia tensor as defined by \Eq{eq:RMIT}.\footnote{Note that several studies also use the non-reduced moment of inertia tensor that is missing the $1/r^2$ weighting.} 

But before studying and quantifying their orientations we consider it important to clearly define the three types of alignments examined here. Therefore, to better visualize the situation and the actual angles to be investigated, respectively, we prepared \Fig{fig:angles}. The sketch shows the example of a sub-halo residing inside a larger host halo. However, only the relative distance of the subhalo is of relevance for our work and -- as mentioned before -- we also consider situations where the subhaloes are found at distances out to $3\times\Rcrit$. The relevant vectors in the study of alignments are
\begin{itemize}
    \item $\vvec{e}_{a}^{h}$: the major axis $a$ as given by the eigenvector of the reduced moment of inertia tensor of the whole halo corresponding to the smallest eigenvalue,
    \item $\vvec{e}_{a}^{*}$: the major axis $a$ as given by the eigenvector of the reduced moment of inertia tensor of only the stellar component corresponding to the smallest eigenvalue, and
    \item $\vvec{r}$: the position vector of the sub-halo in the rest-frame of the host halo.
\end{itemize}

\noindent
The alignment angles -- that without loss of information are considered to be only in the range $\theta \in [0,\pi/2]$ -- are then defined as follows
%\begin{equation} \label{eq:misalignment}
%    \begin{array}{cc}
%         \theta_{r-h} =  & \displaystyle \arccos{\left|\frac{\vvec{r}\cdot\vvec{e}_{a}^{h}}{|\vvec{r}| |\vvec{e}_{a}^{h}|}\right|},\\
%         \theta_{r-*} =  & \displaystyle \arccos{\left|\frac{\vvec{r}\cdot\vvec{e}_{a}^{*}}{|\vvec{r}| |\vvec{e}_{a}^{*}|}\right|},\\
%         \theta_{h-*} =  & \displaystyle \arccos{\left|\frac{\vvec{e}_{a}^{h}\cdot\vvec{e}_{a}^{*}}{|\vvec{e}_{a}^{h}| |\vvec{e}_{a}^{*}|}\right|}.
%    \end{array}
%\end{equation}
\begin{equation} \label{eq:misalignment}
    \begin{array}{rcl}
         \theta_{r-h} & =  & \arccos{\left|{\hat{\vvec{r}}\cdot\hat{\vvec{e}}_{a}^{h}}\right|},\\
         \theta_{r-*} & =  & \arccos{\left|{\hat{\vvec{r}}\cdot\hat{\vvec{e}}_{a}^{*}}\right|},\\
         \theta_{h-*} & =  & \arccos{\left|{\hat{\vvec{e}}_{a}^{h}\cdot\hat{\vvec{e}}_{a}^{*}}\right|}.
    \end{array}
\end{equation}

 \noindent
 where the $\hat{}$\ -symbol indicates that the vector has been normalized to unity. The first two angles are quantifying `radial' (or `shape') alignment which is the prime objective of our study; the last one measures what we already referred to as `galaxy-halo alignment' and will be briefly investigated, too. Further note that we require $b/a<0.9$ in order to avoid oblate objects $a\sim b>c$ for which alignment with $\vvec{e}_{a}$ cannot be properly measured.

%%%%%%%%%%%%%%%%%%%%%%%%%%%%%%%%%%%%%%%%%%%%%%%%%%
\section{Shapes and their alignments}\label{sec:shapes_and_IA}
%%%%%%%%%%%%%%%%%%%%%%%%%%%%%%%%%%%%%%%%%%%%%%%%%%

\subsection{Shapes} \label{sec:shapes}
%%%%%%%%%%%%%%%%%%%%%%%%%%%%%%%%%%%%%%%%%%%%%%%%%%
\begin{figure*}
% \begin{center}
   \hspace*{-0.1cm}\includegraphics[width=17.5cm]{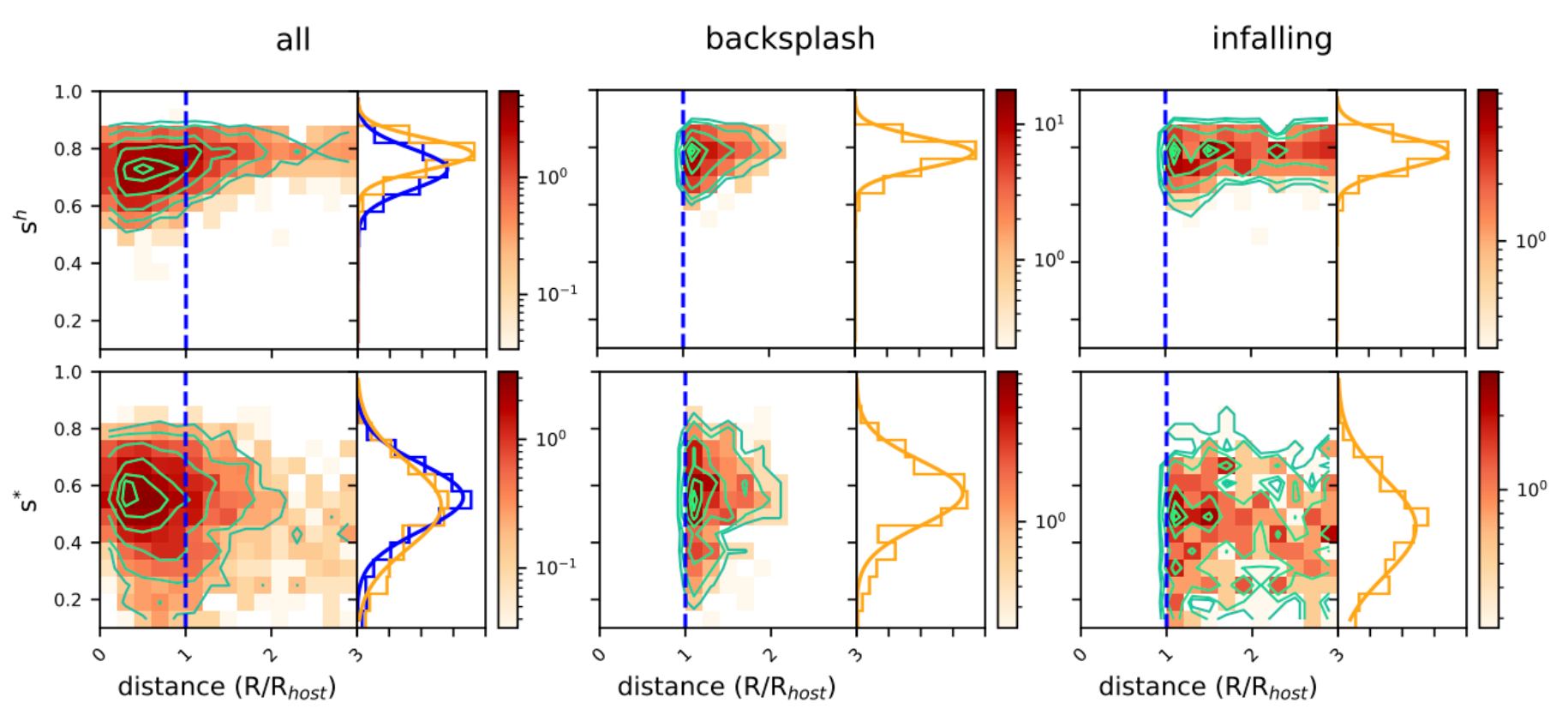}
% \end{center}
   \caption{Halo (upper panel) and stellar component (lower panel) sphericities as a function of distance to the centre of the (host) galaxy cluster. The panels to the right of each contour plot show the respective distribution of sphericities for objects inside (blue) and outside (orange) $\Rhost=\Rcrit$ of the central host cluster. The vertical blue dashed line corresponds to \Rcrit. While we use all objects in the left panel, the data is restricted to backsplash (infalling) objects in the middle (right-most) one. The colour-bar encodes the number of objects in each bin.}
 \label{fig:shapes_distance}
 \end{figure*}

The alignment of the haloes and galaxies can only be analysed in the presence of non-spherical objects. We therefore start with presenting some statistics on the shapes of the sub-haloes and galaxies considered here. While a simple probability distribution could be considered sufficient, we prefer to also take into account the distance of our objects to the central galaxy cluster and  their affiliation to one of our three object samples: subhalo, infalling, and backsplash, respectively. In \Fig{fig:shapes_distance} we therefore show the sphericity for haloes (upper panel) and `galaxies'\footnote{We loosely refer to the stellar component of our selected objects as `galaxy' (cf. \Sec{sec:objectselection}).} (lower panel) in relation to the normalized distance $R/\Rhost$ to the centre of the central galaxy cluster with radius $\Rhost\equiv\Rcrit$. While the left column shows the shape distributions for all our objects, the other columns distinguish between backsplash (middle column) and infalling (right column) objects in order to check for possible effects that transits have on the shape of the different components under study. When comparing haloes and galaxies we observe that a) all our objects are in fact non-spherical (as can be verified by the shape distributions given to the right of each contour plot), b) the halo component follows a more spherical distribution than the galaxy one, and c) the trends with distance are different, i.e. haloes become less spherical when located closer to the central galaxy cluster whereas galaxies show no apparent correlation with distance; this trend is in agreement with results reported for other hydrodynamical simulations \citep[e.g., Fig.8 in][]{Tenneti2015} and will not be quantified in more detail here. Instead we contrast the results for subhaloes to those located \textit{outside} of $\Rhost$. Following previous works \citep[e.g.][]{Ryden2004,Knebe08,Padilla2008}, we fit the shape distributions to a simple Gaussian 
\begin{equation}
    p(s) \propto e^{(s-\mu)^2/2\sigma^2}
\end{equation}
where $\mu$ measures the peak position (i.e. mean) and $\sigma$ the width (i.e. standard deviation). The best-fit parameters are listed in \Tab{tab:Gaussian_fit_shapes}. The decreasing sphericity of haloes closer to the central galaxy cluster -- as quantified by the distribution shifting towards lower $s^h$ values and the marginal drop in peak position $\mu^h$, respectively -- can be attributed to tidal interactions with the central galaxy cluster and the resulting tails of stripped material that will distort the shape (as some of those particles might still be bound to the subhalo). This is actually a trend opposite to the one found by \citet{Kuhlen07}, but in that work the shape has been measured at $R_{\rm max}$ (i.e. the position of the peak of the circular rotation curve and hence much further in than in our case, cf. \Sec{sec:shapecalculation}) and their simulations are dark matter only. \citet{Kuhlen07} actually expected the trend to be as the one found by us, and attribute their results to measuring the shape deep inside the halo. However, in contrast \citet{Vera-Ciro2014} found no distance relation for (sub-)halo shapes when either measuring $s^h$ at the halo's radius or at $R_{\rm max}$. The stellar component is still shielded by the halo and not as affected by tides as in the case of the halo -- as witnessed by the smaller difference between the corresponding $\mu^*$ values in the two different radial ranges. But it is interesting to note that both the infalling and the backsplash haloes appear to have very similar shape distributions which is different to the one for the subhaloes. This is yet another indication of tidal effects acting on the subhaloes, but not (anymore) on the infalling (backsplash) haloes. The situation is, however, different for the stellar component: here the backsplash galaxies show a similar shape distribution to the satellites inside \Rhost, both being more spherical than the infalling objects. This could be a attributed to tidal heating (during pericentre passage) of the stars leading to a more spherical configuration which is not `re-shaping' and `re-adjusting', respectively, like the halo when leaving the host cluster.  

Even though we do not explicitly show it here, we confirm that we observe the same trends when investigating $q=b/a$, i.e. the ratio between semi- and major-axis of the objects. However, the average $q$ values are approximately $+0.1$ ($+0.2$) larger for the haloes (stellar component).

 \begin{table}
 \caption{Mean and standard deviation values of the distribution of stellar ($\mu^{*}$, $\sigma^{*}$) and dark matter ($\mu^{h}$, $\sigma^{h}$) shapes for our different haloes population.}
	\label{tab:Gaussian_fit_shapes}\vspace{-0.2cm}
	\setlength{\tabcolsep}{7.0pt}
		\begin{center}
			\begin{tabular}{|l|c|cccc|}
			\hline
			 Object sample & distance & $\mu^{*}$  &   $\mu^{h}$ &   $\sigma^{*}$    & $\sigma^{h}$  \\
			\hline
			All haloes        & r $< \Rcrit$    &   0.56   &     0.73     &     0.12    &  0.07  \\
			                  & r $\geq \Rcrit$ &   0.53   &     0.78     &     0.15    &  0.06  \\
			\hline
			Backsplash haloes & r $\geq \Rcrit$ &   0.58   &     0.78     &     0.14    &   0.05 \\
			\hline
			Infalling haloes  & r $\geq \Rcrit$ &   0.46   &    0.78      &     0.17    &   0.06 \\
			\hline
			\end{tabular}
		\end{center}
	\end{table}

\begin{figure}
   \hspace*{-0.1cm}\includegraphics[width=8.5cm]{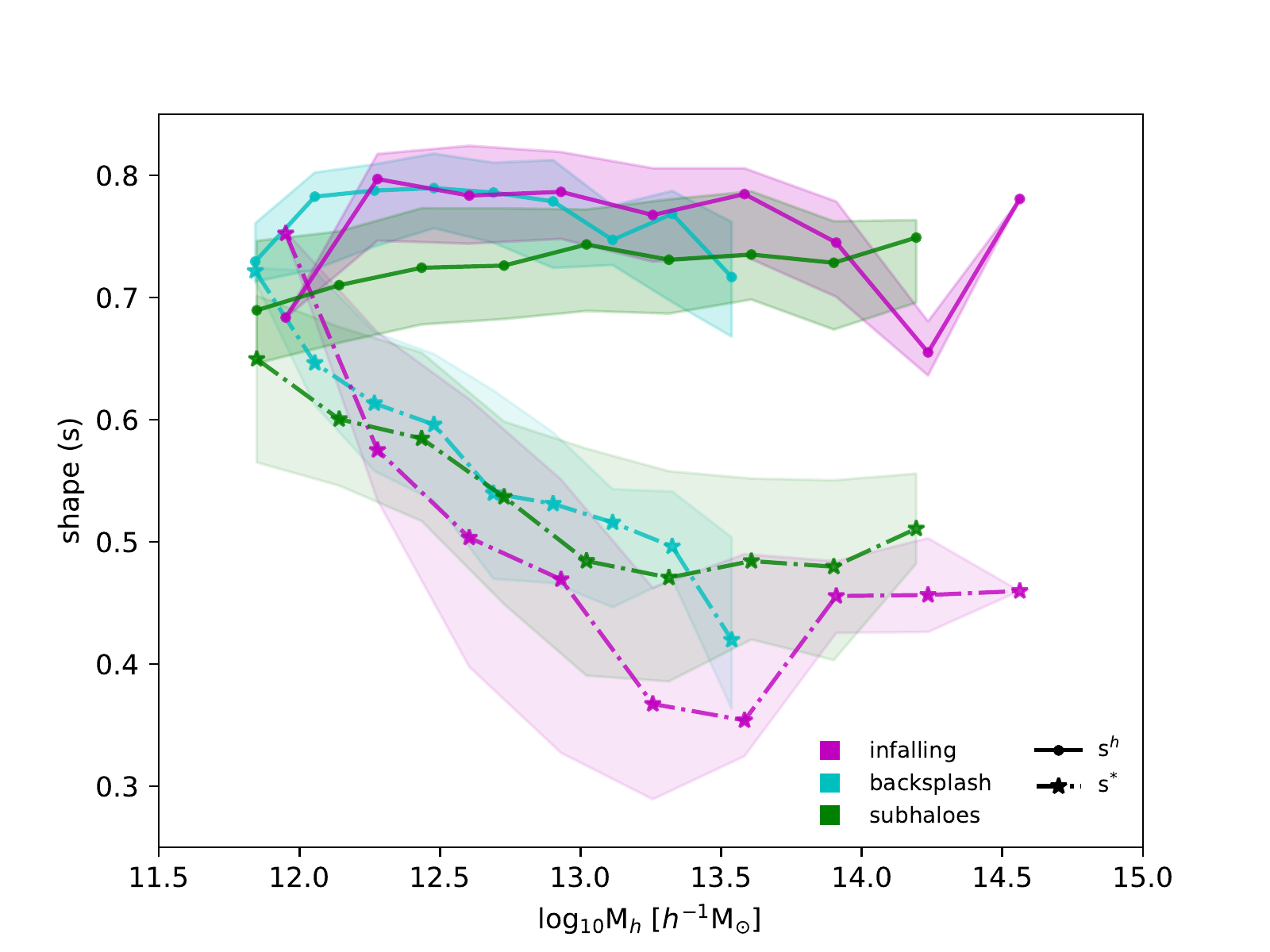}
   \caption{Halo and stellar component sphericities as a function of halo mass. In this and all subsequent plots, the values shown are the median in the respective bin with error bars representing 25/75 percentiles.}
\label{fig:shape_mass}
\end{figure}

There is quite a body of theoretical work in connection to the relation between the shape and halo mass including field haloes \citep[][]{Allgood06, Hahn07, Chisari2017, Chua2019}, subhaloes \citep[][]{Kuhlen07,Vera-Ciro2014,Tenneti14, Barber2015, Velliscig15a,Bhowmick2019}, and also the stellar component \citep[][]{Tenneti14,Tenneti2015b, Bhowmick2019}. They all report a mild dependence of shape on halo mass in the sense that more massive haloes appear to be less spherical; \citet{Tenneti14} even provide a fitting function for this relation. We also investigate such a correlation in \Fig{fig:shape_mass} where we again separate our sample of objects into infalling, backsplash, and sub-haloes (lines are medians with 25/75 percentiles error bars shown as transparent regions). While we confirm a mild trend for infalling and backsplash haloes, it appears that our subhalo shapes are independent of mass. To understand this we have to bear several factors in mind. First and foremost -- and as mentioned before in \Sec{sec:shapecalculation} -- our shapes are biased towards larger sphericities given the nature of our halo finder. When applying the post-correction suggested by \citet{Bailin05} we recover a correlation between shape and mass though still not as strong as reported in previous works and as given by the fitting function provided by \citet{Tenneti14}. But we also need to bear in mind that the definition of the radius \Rcrit\ of our central galaxy cluster that is used to define subhaloes is smaller than the virial radius by approximately a factor $\sim 1.4$ (assuming a NFW profile for the central galaxy cluster with concentration $c\sim 4$). When extending the definition of subhaloes to also include objects out to $1.4\Rcrit$ as well as post-correcting our sphericities as $s\rightarrow s^{\sqrt{3}}$ we do in fact recover the relation advocated by \citet{Tenneti14} (see \Fig{fig:ShapeMassCorrected} \App{app:supplements}). When viewing the mass-distance relation for our objects (not shown here) we attribute this to an increase of halo mass with distance, i.e. the haloes in the outskirts of the central galaxy cluster are on average more massive than those found within \Rcrit\ and hence contribute with their less spherical shapes to the relation when measured at distances $\geq\Rcrit$.

The shape of the stellar component clearly depends inversely on halo mass, with a tendency for no correlation above $M_h>10^{13.5}\hMsun$. We attribute this to the dependence of stellar-to-halo mass on actual halo mass that decreases for increasing halo mass for the objects under consideration here \citep[as naturally expected, e.g. Fig.6 in][]{Cui18}: for higher SMHM values the shape of the stellar component is more in agreement with that of its halo and hence more spherical (something confirmed by plotting the relation between SMHM and stellar shape, also not explicitly shown here). Further, when checking the distance dependence of the SMHM relation (not shown here, too) we find that -- for objects situated close to the centre of the host cluster -- it steeply increases towards our maximum allowed value (cf. \Sec{sec:objectselection}). This is readily explained by tidally truncating the halo making it smaller and more comparable in size to the stellar component. This then leaves the latter `un-shielded' and hence prone to tidal heating. But it does not explain why we find the same (stellar) shape-halo mass trends for the infalling and backsplash population. For the latter two samples we find that neither halo nor stellar mass depend on distance leaving the median stellar-to-halo mass constant. But we observe the same halo mass trend for their SMHM relation and hence conclude that the same explanation applies, i.e. for higher SMHM values the stellar shape more closely follows the halo shape.

\begin{figure}
% \begin{center}
   \hspace*{-0.1cm}\includegraphics[width=8.5cm]{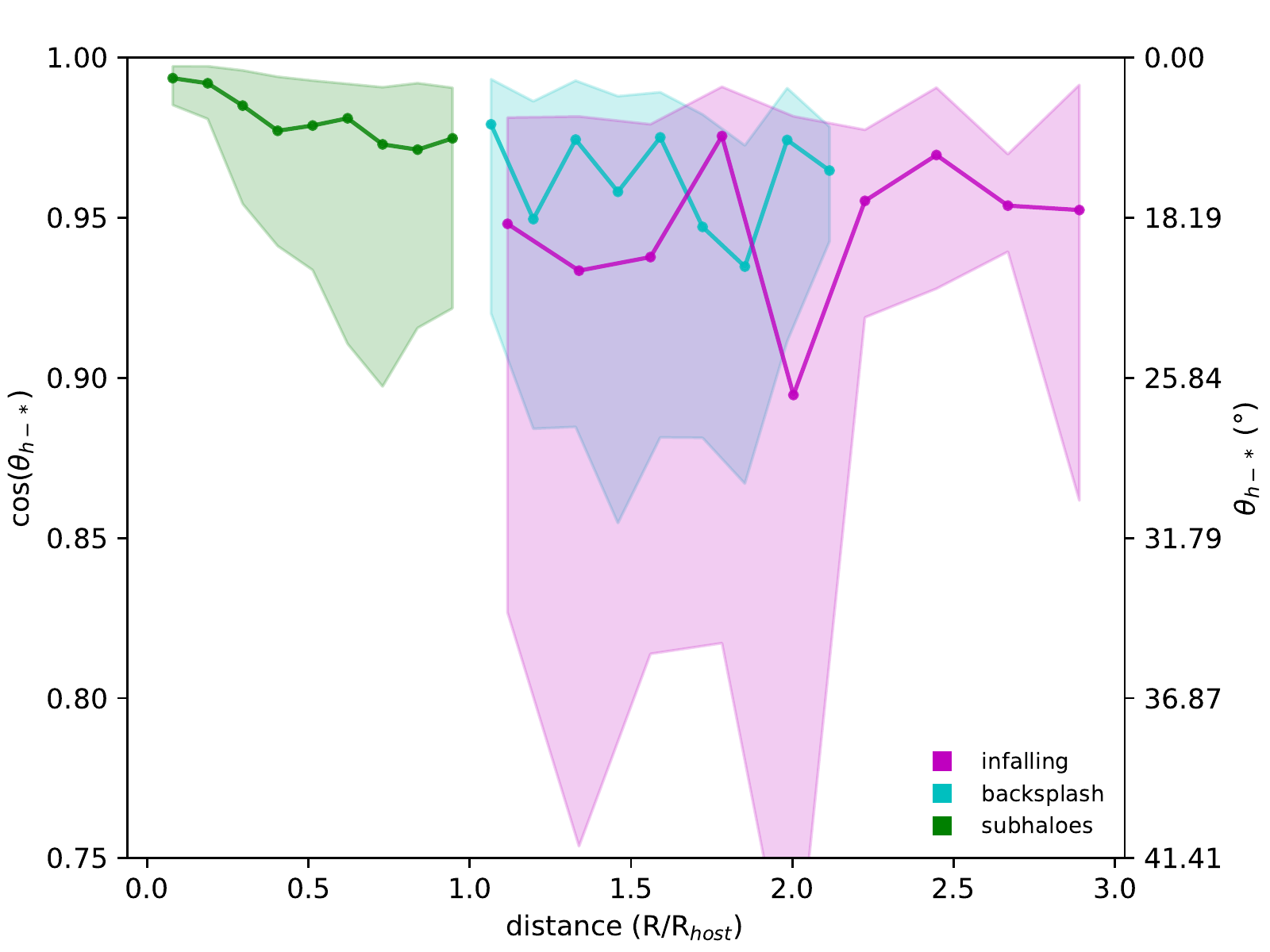}
% \end{center}
   \caption{Median alignment between halo and stellar component as a function of distance to the (host) galaxy cluster; errors are again 25/75 percentiles. Values closer to unity are representative of a stronger alignment.}
 \label{fig:halogalaxyalignment_distance}
 \end{figure}

\subsection{Galaxy-Halo Alignment} \label{sec:IA}
%%%%%%%%%%%%%%%%%%%%%%%%%%%%%%%%%%%%%%%%%%%%%%%%%%
Even though our prime focus lies with the radial alignment, we nevertheless also investigate how the stellar component orients itself with respects to its own host halo as this provides the link between halo and `galaxy' radial alignment. For this purpose, we study the alignment between the major eigenvector of the reduced inertia tensor of the stellar component ($\vvec{e}_{a}^{*}$) and that of the halo they reside in ($\vvec{e}_{a}^{h}$), an alignment referred to as `galaxy-halo' alignment. The result can be viewed in \Fig{fig:halogalaxyalignment_distance} where we show the median $\cos{\theta_{h-*}}$ as a function of distance of the object to the central cluster (errors are again 25 and 75 percentiles). As before, we separate our object sample into the three populations introduced in \Sec{sec:objectselection}. We observe for all populations that the major axis of the stellar component aligns very well with the major axis of its halo, as previously reported in other works \citep{Tenneti14,Tenneti2015,Velliscig15b,Chisari2015,Tenneti2017}. A (mild) radial dependence can only be observed for subhaloes, indicating an even better alignment of stellar and halo shape for galaxies situated closer to the cluster's centre. Such a trend can be explained by the fact that -- due to tidal stripping -- the shape of haloes closer to the centre of the cluster is determined by its innermost part. And -- as has been previously shown \citep[e.g.][]{Bailin05,Velliscig15a} -- the alignment between the stellar and dark matter component increases when considering the central region of objects, i.e. a tidally truncated halo shows stronger galaxy-halo alignment. This improved alignment between the shapes of the stellar and halo component for subhaloes relates back to an artificially increased SMHM ratio. Please refer to a more elaborate discussion of this phenomenon in \App{app:supplements}.

Infalling and backsplash objects show no significant radial dependence. But even though a fair fraction of the backsplash objects entered as deep as 0.1--0.2\Rhost\ into the central cluster \citep[cf. Fig.1 in][]{Haggar2020}, we do not observe that the alignment found for subhaloes is preserved when they exit the host again. This is indicative of differing mechanisms re-adjusting stellar and halo alignment while orbiting, e.g. the halo is more prone to be affected by (and hence re-aligned to) the cluster's potential than the central stellar component.

\begin{figure}
% \begin{center}
   \hspace*{-0.1cm}\includegraphics[width=8.5cm]{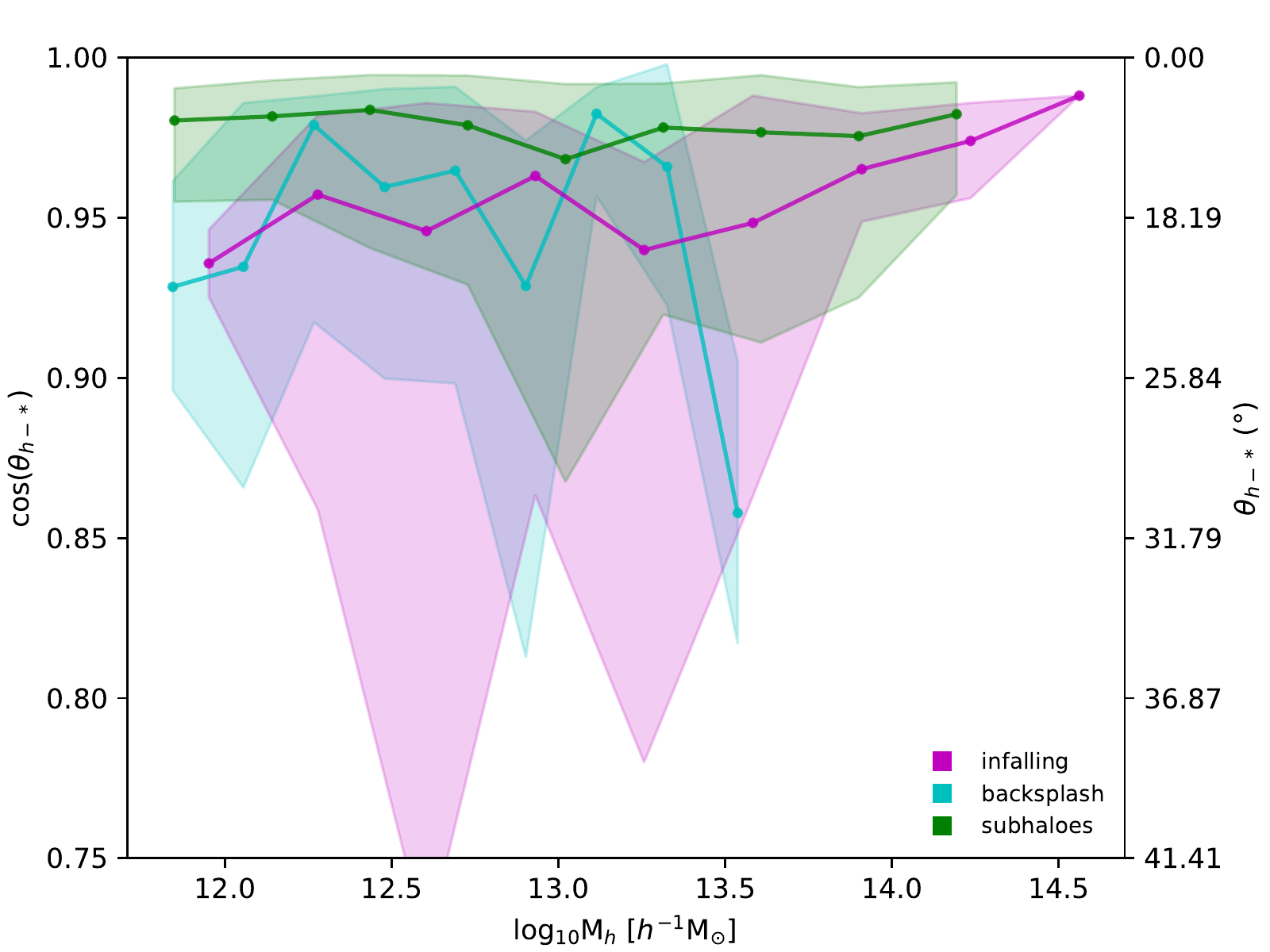}
% \end{center}
   \caption{Alignment between halo and stellar component as a function of halo mass. Values closer to unity are representative of a stronger alignment.}
 \label{fig:halogalaxyalignment_mass}
 \end{figure}

%We can further see -- as shown in \Fig{fig:halogalaxyalignment_mass} -- that the alignment between the stellar component and its halo depends on halo mass, especially for our lower mass objects. \MG{We no longer see this dependency on halo mass. The results obtained for backsplash haloes are irregular.} In the case of both backsplash and infalling objects, the lower mass objects tend to show marginally less alignment forming angles of $\theta_{h-*}>35$\textdegree -- in agreement with previous findings \citep[][]{Tenneti14,Velliscig15a,Chisari2017,Xia2017,Bhowmick2019}. As halo mass increases, the intrinsic alignment of both populations sharply increases up to values of $\theta_{h-*}\lsim26$\textdegree at $M_{\rm halo}\gsim10^{12.5}\hMsun$. Part of the remaining mis-alignment can be attributed to the dependence of shape of halo mass. As we have seen in \Sec{sec:shapes} the stellar component becomes more spherical for lower halo masses and resembles more the shape of the halo, respectively. Therefore, quantifying alignment becomes more difficult. While the same argument should hold for the subhaloes we nevertheless have seen in \Fig{fig:shape_mass} that the subhalo shapes for lower halo masses are marginally less spherical than the corresponding ones for infalling and backsplash objects. We therefore find a much weaker dependence of intrinsic alignment on halo mass for subhaloes.

We can further see -- as shown in \Fig{fig:halogalaxyalignment_mass} -- that the alignment between the stellar component and its halo does not depend on halo mass -- at least not for subhaloes. Previous studies \citep[i.e.][]{Tenneti14,Velliscig15a,Chisari2017,Xia2017,Bhowmick2019} state that lower mass objects tend to show marginally less alignment, as the stellar component becomes more spherical for lower halo masses hence making it more difficult for alignment to be quantified. However, we only observe such a tendency for the infalling population that has not yet experienced any influence of the central galaxy cluster. 

%%%%%%%%%%%%%%%%%%%%%%%%%%%%%%%%%%%%%%%%%%%%%%%%%%
\section{Radial Alignment} \label{sec:radialalignment}
%%%%%%%%%%%%%%%%%%%%%%%%%%%%%%%%%%%%%%%%%%%%%%%%%%
We now turn to the study of radial alignment, i.e. the alignment of the shape of an object with respect to its positional vector in the rest-frame of the central galaxy cluster. Previous analyses of the alignment of the major axis of dark matter (sub-)haloes have shown that the major axis $\vvec{e}_a$ is in fact preferentially oriented towards the centre of the (host) galaxy cluster \citep[e.g.][]{Kuhlen07, Pereira08, Faltenbacher08, Knebe08b, Knebe10a}. However, there are to date only a few studies that made use of full physics hydrodynamical simulations investigating this phenomenon also for `galaxies' \citep[][]{Chisari2017} or even `satellite galaxies' \citep[][]{Knebe10a,Tenneti2015,Velliscig15b,Barber2015}. Further, so far there is no study that distinguishes between infalling and backsplash objects. This work now aims at adding to this by analysing radial alignment in cluster environments and the different factors that may play a role.

\begin{figure}
   \hspace*{-0.1cm}\includegraphics[width=8.5cm]{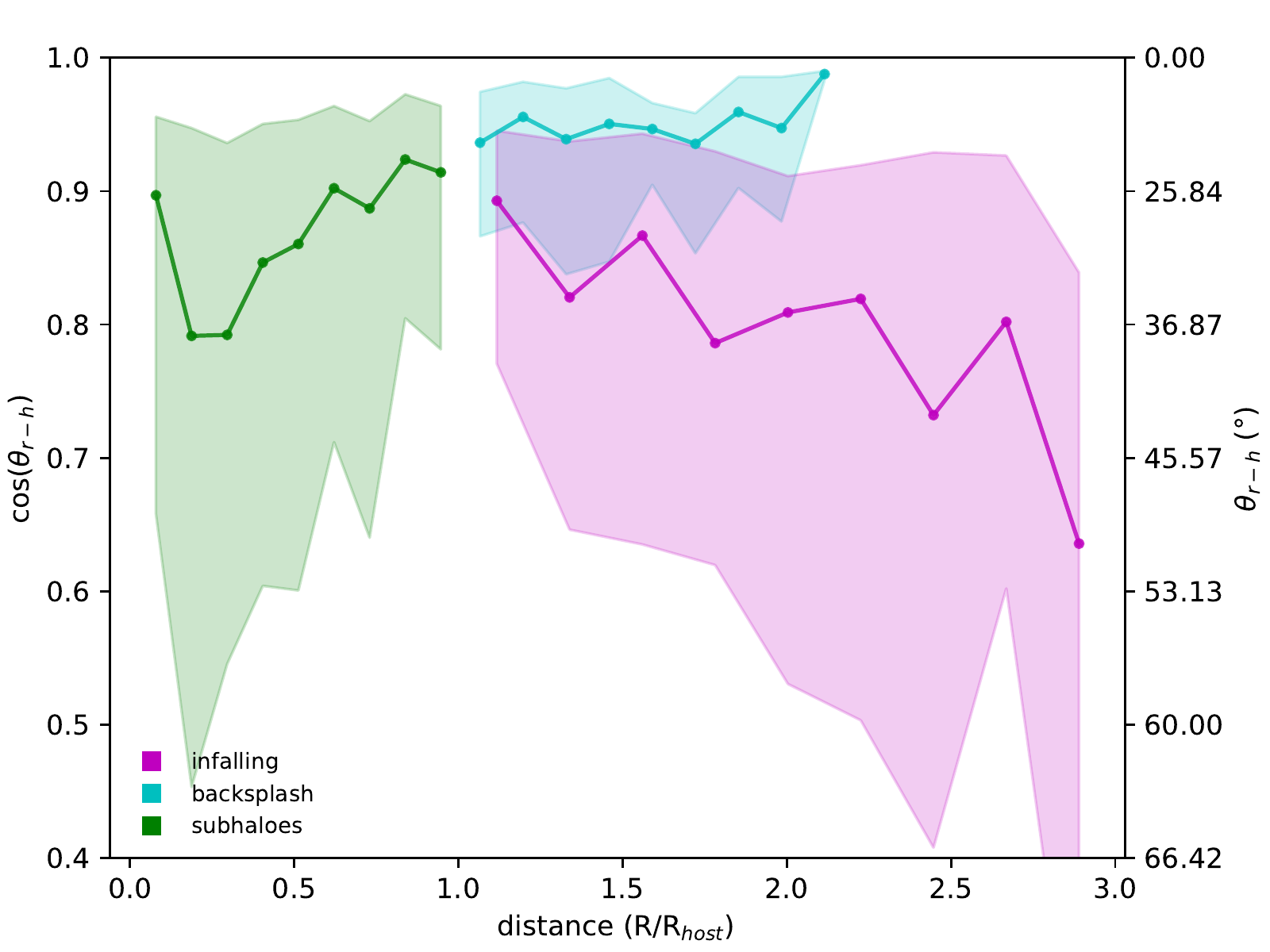}
   \hspace*{-0.1cm}\includegraphics[width=8.5cm]{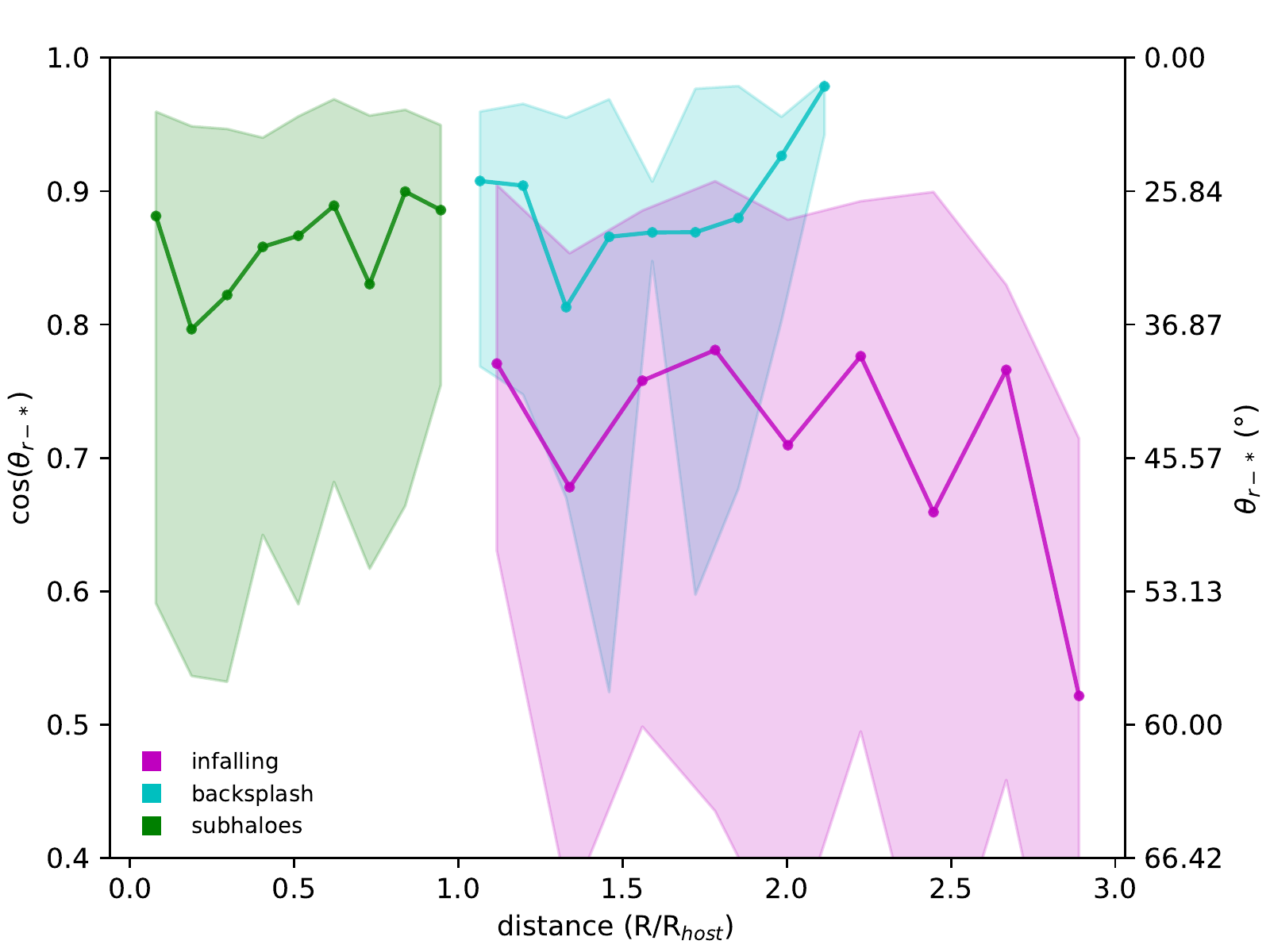}
   \caption{Median radial alignment -- as defined by \Eq{eq:misalignment} using 25/75 percentiles as error bars -- of dark matter (upper panel) and stellar (lower panel) objects as a function of distance to the central galaxy cluster.}
 \label{fig:radialalignment_distance}
 \end{figure}

 \begin{figure}
 % \begin{center}
   \hspace*{-0.1cm}\includegraphics[width=8.5cm]{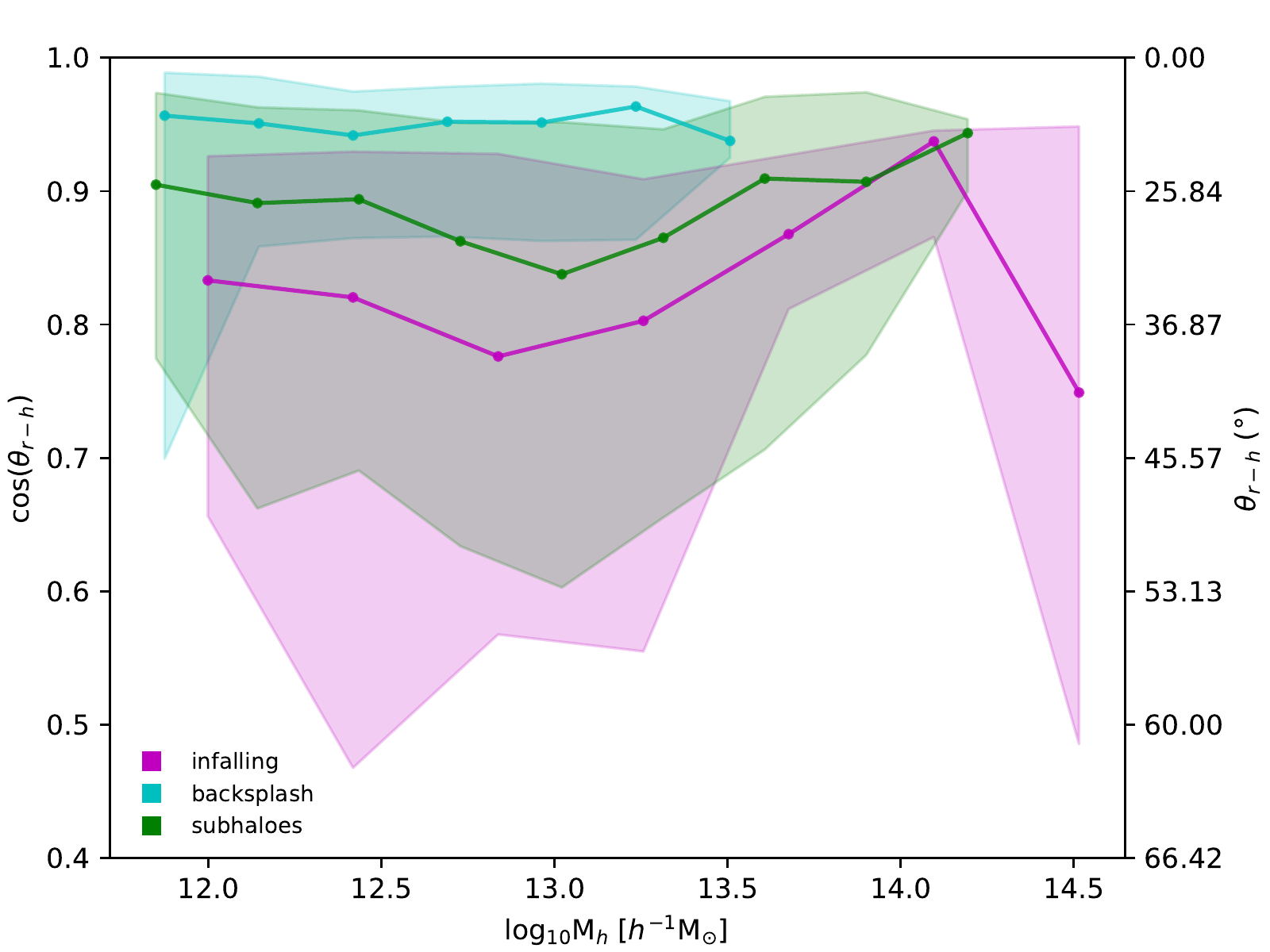}
   \hspace*{-0.1cm}\includegraphics[width=8.5cm]{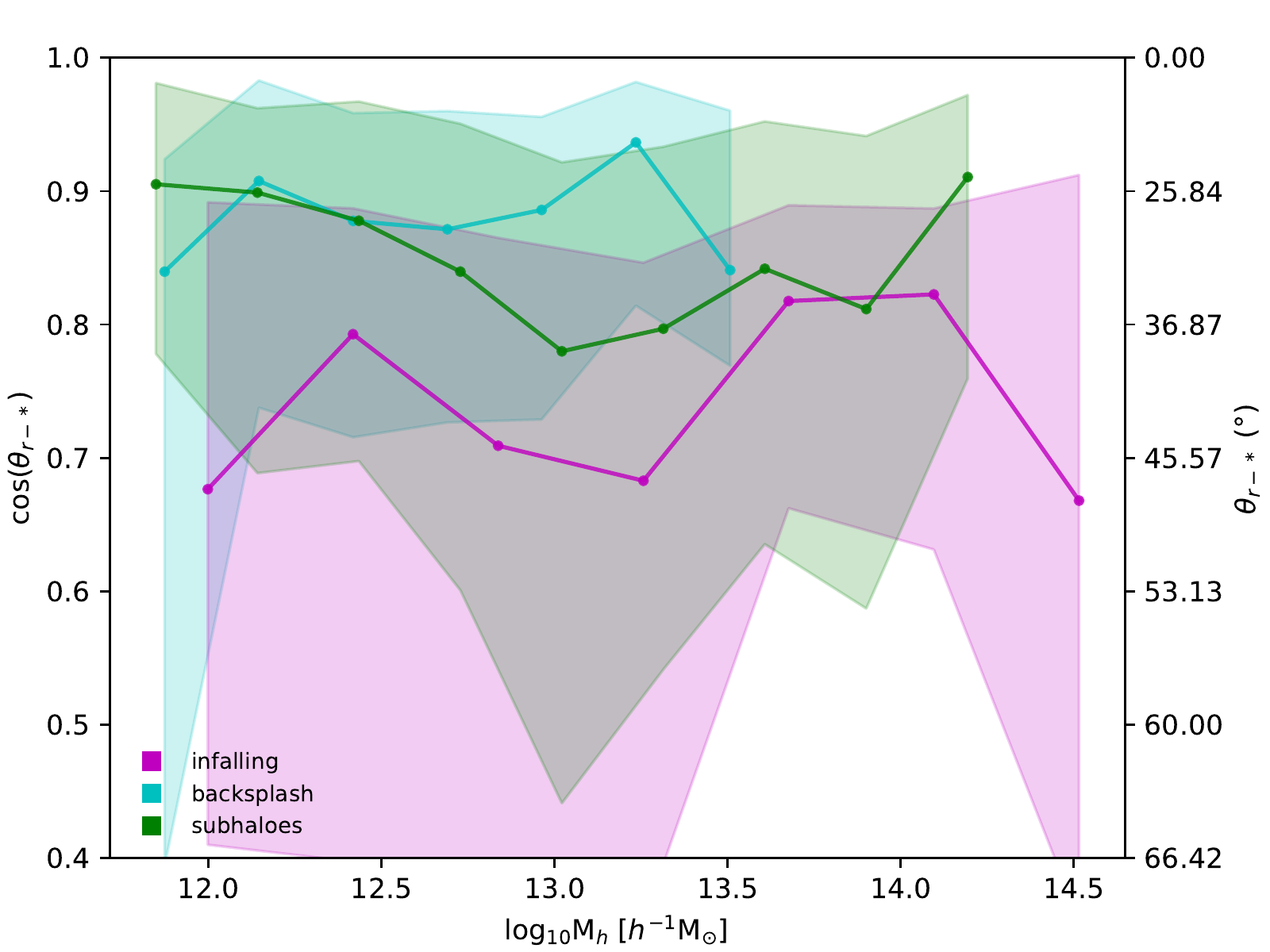}
 % \end{center}
   \caption{Median radial alignment of the dark matter ($\theta_{r-h}$) (upper panel) and stellar ($\theta_{r-*}$) (lower panel)  component of objects as a function of their halo mass (error regions are again 25/75 percentiles).}
  \label{fig:radialalignment_mass}
  \end{figure}

\subsection{Dependence on distance to the centre of the host cluster} \label{sec:RA_distance}
%%%%%%%%%%%%%%%%%%%%%%%%%%%%%%%%%%%%%%%%%%%%%%%%%%
In \Fig{fig:radialalignment_distance}, we show the dependence of radial alignment of dark matter haloes (upper panel) and galaxies (lower panel) on their relative distance to the central galaxy cluster. We clearly see that there is a difference between the radial alignment signals for our three object samples. Regarding subhaloes, we can distinguish -- for both the stellar and dark matter component -- between two different regimes: while they are highly aligned at the edge of the host galaxy cluster (i.e. $R\sim\Rhost$), this alignment tends to decrease as they approach to the host centre, having their smallest value at $R\sim$ 0.2$\Rhost$. This indicates that pre-aligned subhaloes probably have large infall velocities leaving no time for shape readjustment when approaching to the centre \citep[][see also \Fig{fig:velocity_distance} in \App{app:supplements}]{Pereira08,Knebe10a}. However, at distances $R\leq$ 0.2$\Rhost$ we see that radial alignment sharply increases again.\footnote{We have explored and excluded the possibility that this rise is a numerical artifact due to our binning procedure.} This change of trends can be explained by radial vs. circular orbits of subhaloes -- something we will study in \Sec{sec:RA_object_velocity} when quantifying different relations between the bulk velocity $\vvec{v}$ and their positional vector $\vvec{r}$ of our structures. Only subhaloes on radial orbits reach as deep into the central cluster's potential as $R\leq$ 0.2$\Rhost$. If they now have their shapes aligned upon infall this alignment then this persists even at these small distances. A clear indication of the re-adjustment of the subhalo shapes aligning themselves with the positional vector is the rather strong signal found for the backsplash population: the alignment is preserved when exiting the host cluster again.

In the case of infalling objects, we can see that radial alignment for both stellar and dark matter components increases with decreasing distance to the host centre, nearly matching up with the level found for subhaloes at $\Rhost$ -- at least for the halo, not for the stellar component though. This offset in the signal for haloes (upper panel) and galaxies (lower panel) relates back to the marginally larger misalignment between those two components for infalling objects as seen in \Fig{fig:halogalaxyalignment_distance}. At even larger distances from the galaxy cluster, the radial alignment of objects approaching the central cluster tends to decrease even more -- which only appears to be natural. In fact, when extending the study to distances as far as 13\hMpc, i.e. close to the edge of the high-resolution region in which the central galaxy cluster resides, we find that any signal of radial alignment approaches $\cos{\theta}\sim0.5$, as expected for a random distribution. As pointed out by \cite{Pereira08}, the alignment already seen at distances $R\leq\Rhost$ can be a consequence of a primordial alignment of (sub-)haloes with respect to the filaments at which they form \citep[e.g.][]{Ganeshaiah2018}. They are then accreted onto galaxy clusters which lie at the intersection of such filaments only enhancing the aligning by re-adjusting their shape towards the cluster.

Up to now we have treated all our central galaxy clusters equally not distinguishing between relaxed and un-relaxed objects. But as seen by \citet{Mostoghiu18}, such a separation might reveal interesting insights. We therefore also split the sample of galaxy clusters using the same criterion as presented in \citet[][Sec.~3.1.2]{Cui18}. However, we defer from showing the results here as we cannot confirm any clear relation to the dynamical state of the galaxy cluster: neither have the trends already seen in \Fig{fig:radialalignment_distance} changed nor does the actual strength of the signal differ for relaxed and un-relaxed host clusters. This might be attributed to the fact that relaxation is temporary and changes quickly. But we also need to remember that we infer the cluster dynamical state mainly by its dark matter component, see indicators in \citet{Cui18}, and these are not strictly related to the galaxy distribution.

 \subsection{Dependence on halo mass}
% %%%%%%%%%%%%%%%%%%%%%%%%%%%%%%%%%%%%%%%%%%%%%%%%%%
We have seen in \Fig{fig:halogalaxyalignment_mass} that the alignment between the shape of the stellar component and its halo shows no prominent correlation with halo mass -- at least not for subhaloes and backsplash galaxies. We now confirm in \Fig{fig:radialalignment_mass} that radial alignment as a function of halo mass shows in general but weak trends: for subhaloes we might infer a minute increase of alignment for objects with $M_h<10^{13}\hMsun$.  We therefore conclude that the mass of the object only plays a minor role in observations of radial alignment, in agreement with \citet[][right panel of Fig.~14]{Tenneti2015b}. Further, splitting the host galaxy clusters into relaxed and un-relaxed systems \citep[using the criteria specified in][]{Cui2017,Cui18} does not reveal any relation to the dynamical state of the host and hence we decided to omit such a plot for clarity. 

\begin{figure}
   \hspace*{-0.1cm}\includegraphics[width=8.5cm]{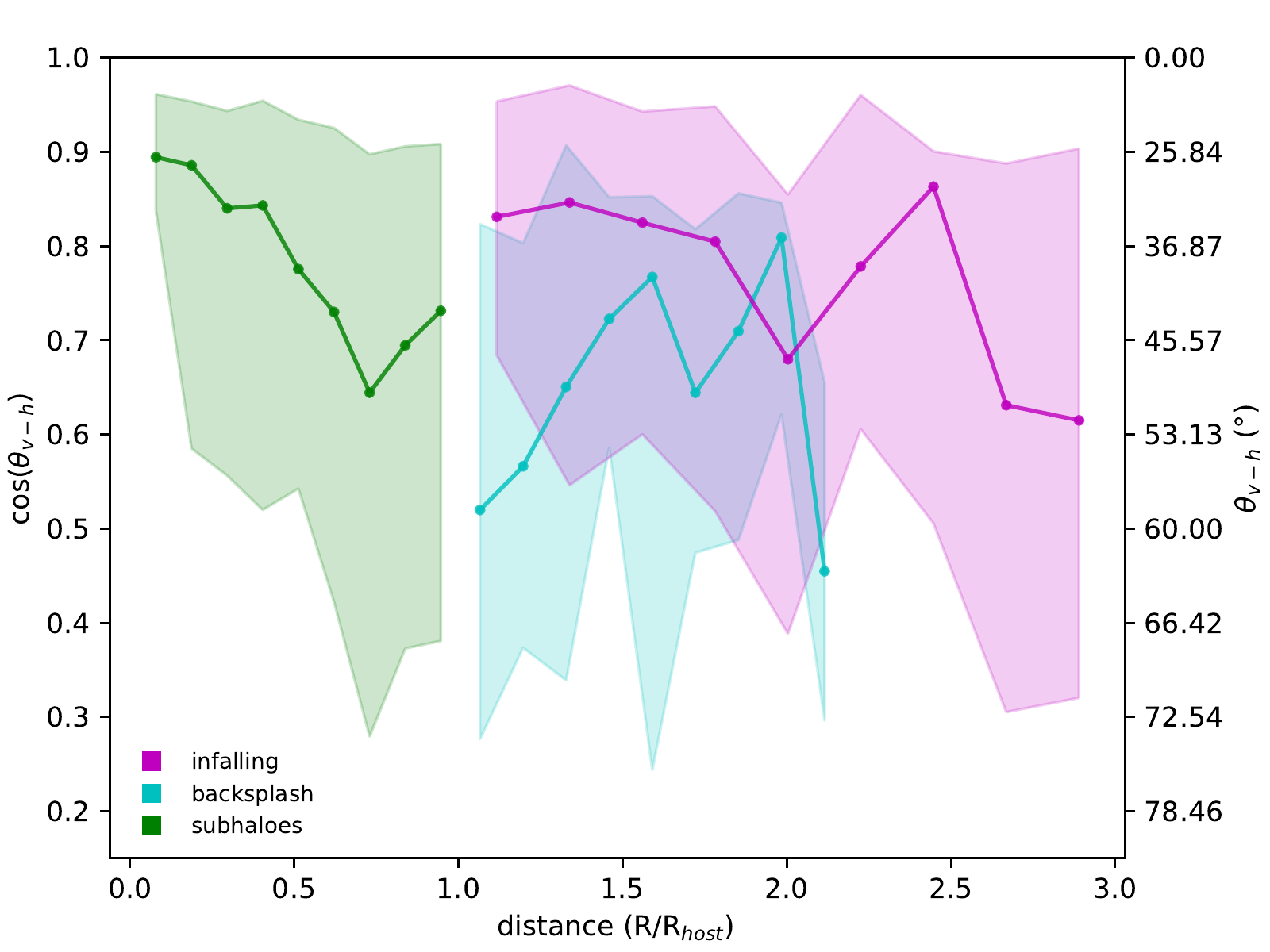}
   \hspace*{-0.1cm}\includegraphics[width=8.5cm]{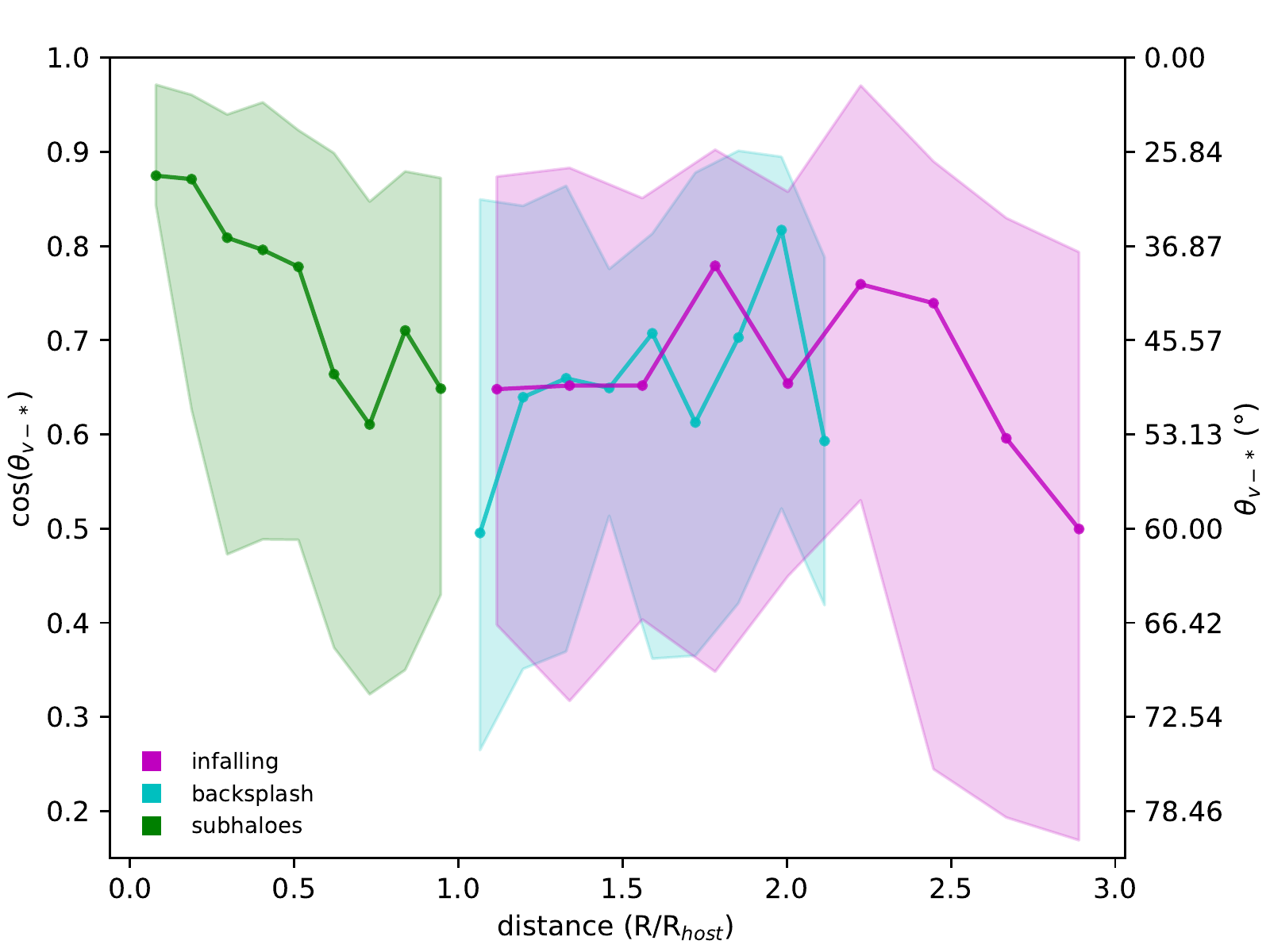}
   \caption{Alignment of the halo (upper panel) and stellar (lower panel) component with the bulk velocity vector of the halo in the rest frame of the central galaxy cluster. Values closer to unity are representative of the object's major axis $\vvec{e}_a$ being aligned with the direction in which the object is moving.}
 \label{fig:flowalignment_distance}
 \end{figure}

\subsection{Relation to object velocity}\label{sec:RA_object_velocity}
\begin{figure}
% \begin{center}
   \hspace*{-0.1cm}\includegraphics[width=8.5cm]{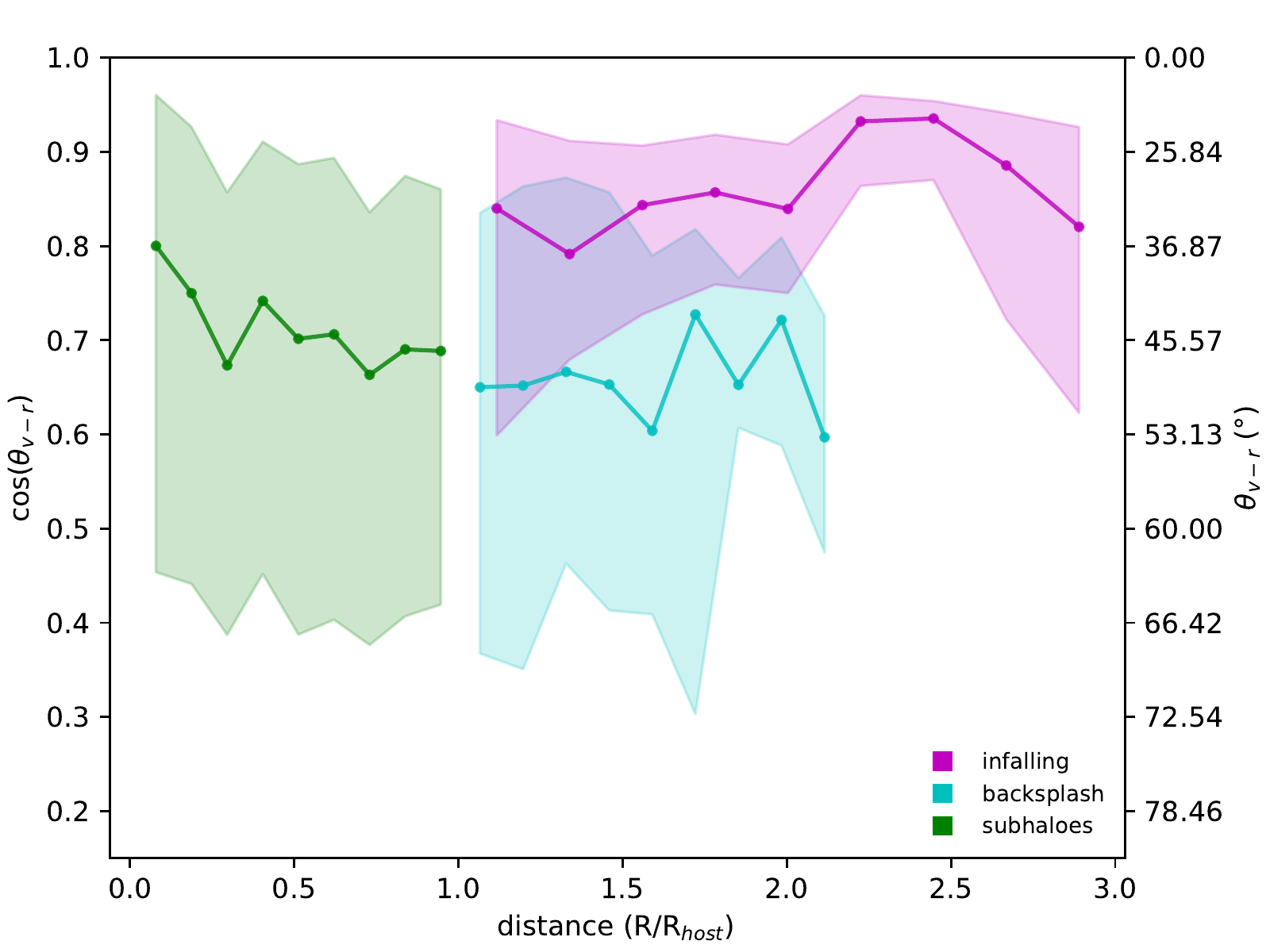}
% \end{center}
   \caption{Alignment between the bulk velocity vector and the radial position vector of the haloes in the rest frame of the central galaxy cluster. Values closer to unity are representative of a stronger alignment.}
 \label{fig:velocity_position_distance}
 \end{figure}

%%%%%%%%%%%%%%%%%%%%%%%%%%%%%%%%%%%%%%%%%%%%%%%%%%
In \Fig{fig:radialalignment_distance} we have seen that -- for both the stellar and halo shapes -- the radial alignment signal for subhaloes weakens for objects in the distance range $R \sim \Rhost \rightarrow 0.2\Rhost$. Previous works \citep[e.g.][]{Faltenbacher08, Knebe10a} attributed this to tidal torquing and the fact that these objects -- presumably on bound orbits -- move too fast at their pericentre for radial alignment to be effective, i.e. the subhalo does not have enough time to re-arrange its shape at pericentre passage. However, we have also seen that the radial alignment strengthens when considering even smaller distances $R < 0.2\Rhost$. But this certainly requires a different explanation. We therefore will have a closer look at the velocities of our objects and show in \Fig{fig:flowalignment_distance} the alignment between the object's bulk velocity vector $\vvec{v}$ (in the rest frame of the central host galaxy cluster) and its major axis $\vvec{e}_{a}^{h/*}$, i.e.
\begin{equation}
    \theta_{v-h/*} =  \arccos{\left|{\hat{\vvec{v}}\cdot\hat{\vvec{e}}_{a}^{h/*}}\right|}  \ ,
\end{equation}
where the $\hat{}$\ -symbol again indicates a normalized vector. For subhaloes, the plot shows a clear tendency for these two vectors to be aligned as they approach to the host centre, both for their halo and stellar components (as expected from \Fig{fig:halogalaxyalignment_distance}). This then certainly explains the drop in radial alignment towards $0.2\Rhost$, but not the subsequent rise again. We conjecture that this rise has to be caused by a population of subhaloes with different orbital properties. In general we expect subhaloes to be on bound orbits with eccentricities $p=1-p/a \sim 0.6$ where $p$ and $a$ are peri- and apo-centre, respectively \citep[cf. Fig.7 in][]{Gill04b}. Hence velocity and position vector will not be aligned: if they were aligned, it would mean that the subhalo is on a highly radial orbit with $p\rightarrow1$. We therefore check in \Fig{fig:velocity_position_distance} for exactly this alignment, i.e. the plot shows 

\begin{equation}
    \theta_{v-r} =  \arccos{\left|{\hat{\vvec{v}}\cdot\hat{\vvec{r}}}\right|}  \ ,
\end{equation}

\noindent
where both $\vvec{v}$ and $\vvec{r}$ are given for the whole halo in the rest frame of the central galaxy cluster. We can clearly see that there is a significant difference between the results obtained for each of our halo populations. Focusing first on the subhaloes, we find that those objects that are closer to the host centre and with distances $R\leq0.2\Rhost$, respectively, show an increase in the velocity-position alignment indicative of more radial orbits. That then explains why we found the increase of radial alignment in the same distance regime: the objects are plunging into the central galaxy cluster with their major axis $\vvec{e}_a$ aligned with the flight path. The same can be observed for the infalling haloes further supporting the scenario in which radial alignment of haloes at large distances is a consequence of the accretion onto galaxy clusters through the large-scale structure: they are on highly radial orbits yet gradually developing a radial alignment of their shape (as observed in \Fig{fig:radialalignment_distance}). But subhaloes in the radial range $R\in [0.2,1.0]\Rhost$ show a weaker velocity-position alignment which is then due to moving on more regular orbits with $p\sim0.6$. We finally notice that backsplash haloes show the least velocity-position alignment more or less matching the signal for subhaloes at \Rhost\ -- in agreement with the results found for the radial alignment in \Fig{fig:radialalignment_distance}: the backsplash haloes are keeping their orientations pointing towards the host centre while being on trajectories closer to circular than the infalling population. This significant difference between the alignment signals of infalling and backsplash (sub-)haloes makes this analysis useful for separating them.

We close this sub-section with the remark that when dividing our host clusters into relaxed and un-relaxed central galaxy clusters we find that our results are once more independent of the dynamical state of the host cluster.

%%%%%%%%%%%%%%%%%%%%%%%%%%%%%%%%%%%%%%%%%%%%%%%%%%
\section{Conclusions} \label{sec:conclusions}
%%%%%%%%%%%%%%%%%%%%%%%%%%%%%%%%%%%%%%%%%%%%%%%%%%
Using the mass-complete sample of 324 numerically modelled galaxy clusters provided by `The Three Hundred' collaboration\footnote{\url{http://www.the300-project.org}} we investigate how the shape of haloes and galaxies orbiting in and about the central cluster orients itself with respects to position. To this extent we use the stellar component of a halo as a proxy for `galaxy'. We further separate our objects of interest into three distinct populations: a) subhaloes, b) infalling, and c) backsplash objects. After applying very conservative selection criteria to the general pool of objects aiming at minimizing numerical artifacts, we find for the shapes that

\begin{itemize}
    \item backsplash and infalling \textit{haloes} have similar shapes which are more spherical than those of subhaloes, and that
    \item backsplash and satellite \textit{galaxies} have similar shapes which are more spherical than those of infalling galaxies.
\end{itemize}
    
Regarding the radial alignment as measured by the angle between the eigenvector $\vvec{e}_a$ corresponding to the largest eigenvalue of the (reduced) moment of inertia tensor and the position vector in the rest from of the central galaxy cluster we find
    
\begin{itemize}
    \item stronger radial alignment for backsplash than infalling haloes,
    \item equal radial alignment for backsplash and subhaloes at \Rcrit,
    \item radial alignment of subhaloes decreases towards the centre due to the faster movement at pericentre and hence lack of time for shape re-adjustment, but
    \item radial alignment actually increases in the very central regions of galaxy clusters $R\leq0.2\Rhost$ which we attribute to radially aligned objects infalling on highly radial orbits, 
    \item halo and galaxy radial alignment trends are following each other due to a strong alignment of both halo and galaxy shape, and eventually that
    \item radial alignment does not depend on halo mass.
\end{itemize}

When additionally dividing our sample of central galaxy clusters into relaxed and un-relaxed objects (following the same criterion as presented in \citet{Cui18}) we cannot find any influence of the dynamical state on the aforementioned results. This likely relates to the fact that relaxation is temporary and changes quickly.

Both the differences between radial alignment (\Fig{fig:radialalignment_distance}) and velocity-position alignment (\Fig{fig:velocity_position_distance}) for backsplash and infalling objects can serve as a means to actually distinguish these two distinct populations. However, we need to bear in mind that the present analysis has been performed in 3D and the applicability of this finding in an observer's frame in 2D remains to be shown. We leave a projection of the simulation data into an observer's plane to a future study where we will eventually also utilize galaxy colours. Further, while the present study is only based upon present-day data, we will extend it to higher redshifts in a follow-up work. There we will then also trace back the objects quantifying the change in radial alignment of individual objects and investigate possible links to the cosmic web in the vicinity of the central galaxy cluster.

%%%%%%%%%%%%%%%%%%%%%%%%%%%%%%%%%%%%%%%%%%%%%%%%%%
\section*{Acknowledgements}
%%%%%%%%%%%%%%%%%%%%%%%%%%%%%%%%%%%%%%%%%%%%%%%%%%
% Entry for the table of contents, for this guide only
\addcontentsline{toc}{section}{Acknowledgements}
This work has been made possible by the `The Three Hundred' collaboration.\footnote{\url{https://www.the300-project.org}} The project has received financial support from the European Union's Horizon 2020 Research and Innovation programme under the Marie Sklodowskaw-Curie grant agreement number 734374, i.e. the LACEGAL project.

The authors would like to thank The Red Espa\~{n}ola de Supercomputaci\'{o}n for granting us computing time at the MareNostrum Supercomputer of the BSC-CNS where most of the cluster simulations have been performed. Part of the computations with \textsc{GADGET-X} have also been performed at the `Leibniz-Rechenzentrum' with CPU time assigned to the Project `pr83li'.

AK and RM are supported by the MICIU/FEDER through grant number PGC2018-094975-C21. AK further acknowledges support from the Spanish Red Consolider MultiDark FPA2017-90566-REDC and thanks Manfred H\"ubler and Siegfried Schwab for the lions and the cucumber. WC acknowledge supported from the European Research Council under grant number 670193. RH acknowledges support from STFC through a studentship. MDP is supported by funding from Sapienza Universit\`{a} di Roma -- Progetti di Ricerca Medi 2019, prot. RM11916B7540DD8D. KH was supported by SNF Sinergia CRSII5\_173716. CP acknowledges support of the Australian Research Council Centre of Excellence for All Sky Astrophysics in 3D (ASTRO3D), CE170100013.

%%%%%%%%%%%%%%%%%%%%%%%%%%%%%%%%%%%%%%%%%%%%%%%%%%


\clearpage
\bibliographystyle{mnras}
\bibliography{archive}

%%%%%%%%%%%%%%%%%%%%%%%%%%%%%%%%%%%%%%%%%%%%%%%%%%%
\appendix
%%%%%%%%%%%%%%%%%%%%%%%%%%%%%%%%%%%%%%%%%%%%%%%%%%%
In this Appendix we provide complementary plots that support some of the statements made in the main text.

%%%%%%%%%%%%%%%%%%%%%%%%%%%%%%%%%%%%%%%%%%%%%%%%%%%
\section{Supplementary Plots} \label{app:supplements}
%%%%%%%%%%%%%%%%%%%%%%%%%%%%%%%%%%%%%%%%%%%%%%%%%%%
 \begin{figure}
 % \begin{center}
   \hspace*{-0.1cm}\includegraphics[width=8.5cm]{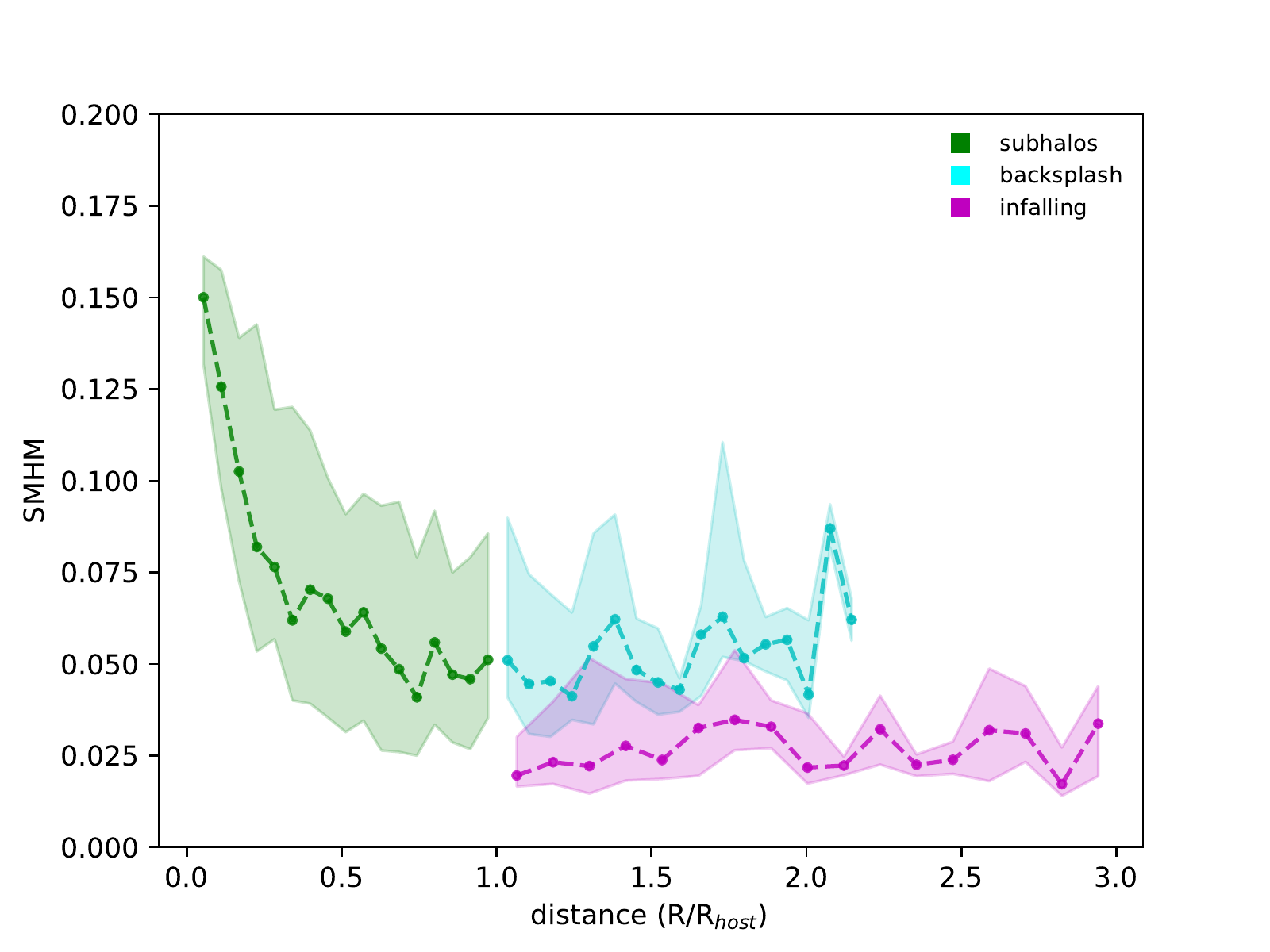}
 % \end{center}
   \caption{SMHM ratio as a function of normalized distance to the host.}
  \label{fig:SMHM_distance}
  \end{figure}

 \begin{figure}
 % \begin{center}
   \hspace*{-0.1cm}\includegraphics[width=8.5cm]{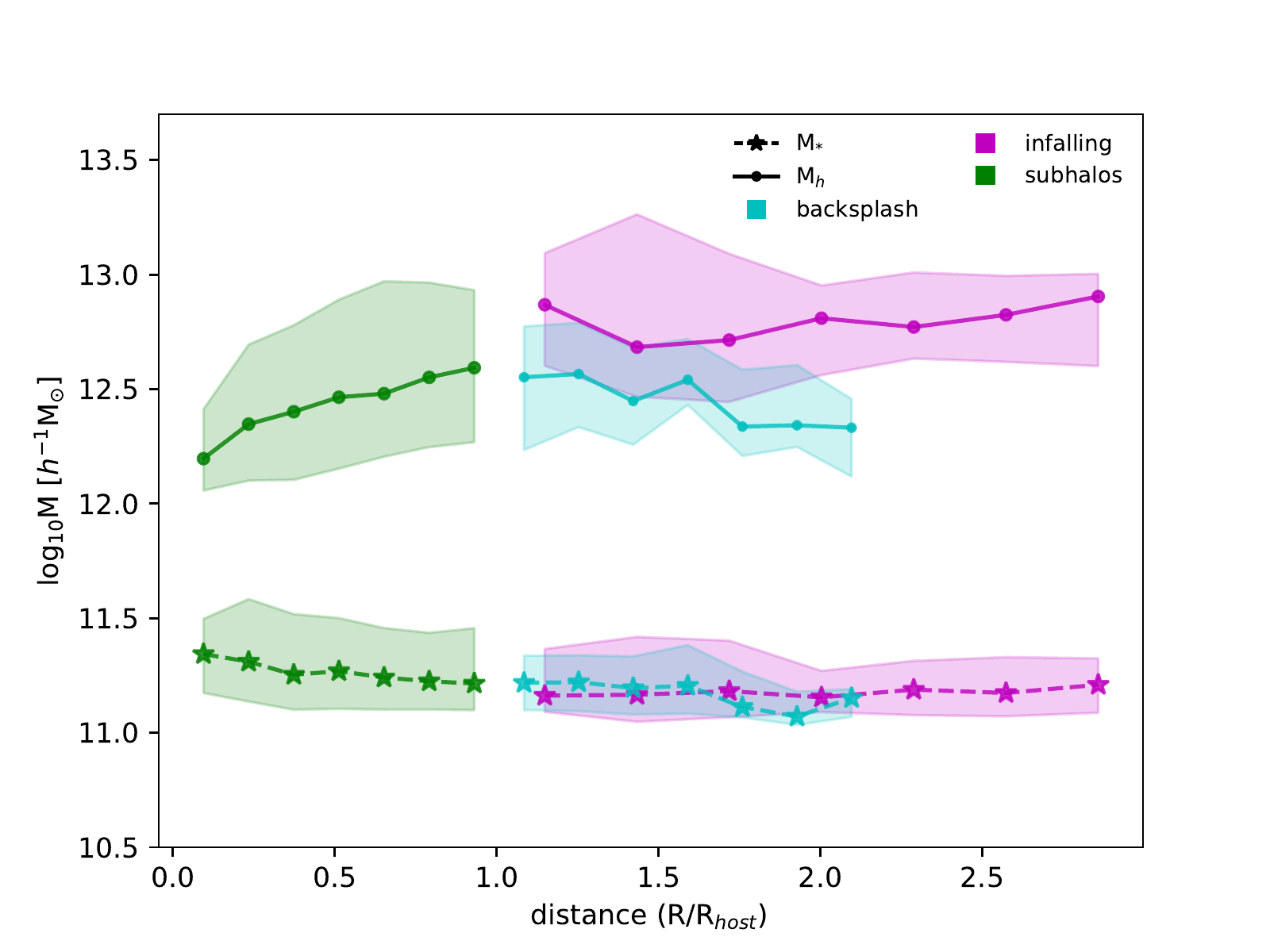}
 % \end{center}
   \caption{Median halo mass as a function of normalized distance to the host.}
  \label{fig:MedMass_distance}
  \end{figure}

We like to return to the point of what causes the rise in alignment between galaxy and halo shape as seen for objects closer to the position of the central galaxy cluster (cf. \Fig{fig:halogalaxyalignment_distance}). If dark matter has been stripped off, then the increased alignment observed for distances $\leq 0.5\Rcrit$ should be preserved for backsplash galaxies. But for those objects we find again an alignment in agreement with subhaloes located at $\sim \Rcrit$. We speculate that the apparent stripping of material for haloes $\leq 0.5\Rcrit$ is a mere numerical artifact. To shed light into this we have present in \Fig{fig:SMHM_distance} the stellar mass to halo mass ratio (SMHM) as a function of distance. We can clearly see how the SMHM ratio increases towards the centre, and that it is continuous at \Rcrit\ for subhaloes and backsplash, as expected. We accompany this plot \Fig{fig:MedMass_distance} that shows the median mass as a function of distance. Both plots taken together suggest that while the stellar component of objects does not depend on distance, the total mass gets smaller for objects closer to the position of the central galaxy cluster.\footnote{We have confirmed that this is not a `volume effect': the mass function of subhaloes indicates that there are far more low mass than high mass haloes and hence we preferentially expect to find low mass haloes in small volumes such as the spheres closer to the host centre. But when randomly assigning masses to the objects entering \Fig{fig:MedMass_distance} we find a flat radial distribution and hence any trend seen in it cannot be driven by such an effect.} But when the objects leave the deep potential well (i.e. backsplash galaxies) the halo mass rises again. This phenomenon - that surely is of numerical nature - has been discussed in great detail in, for instance, \citet[][Fig.7]{Knebe11}, \citet[][Fig.7]{Onions12}, and \citet[][Fig.11]{Knebe13b} where it also has been shown to primarily affect subhaloes $<0.2 \Rcrit$. We also observe above that the SMHM is lowest for infalling due to lack of processing. And we attribute the erratic behaviour for backsplash objects at the largest distances they can have to small number statistics: there are a mere 6 objects in the last two bins. However, we also note that those farthest backsplash galaxies have the lower masses than the other backsplash objects indicative of actual stripping.\\

\begin{figure}
 % \begin{center}
   \hspace*{-0.1cm}\includegraphics[width=8.5cm]{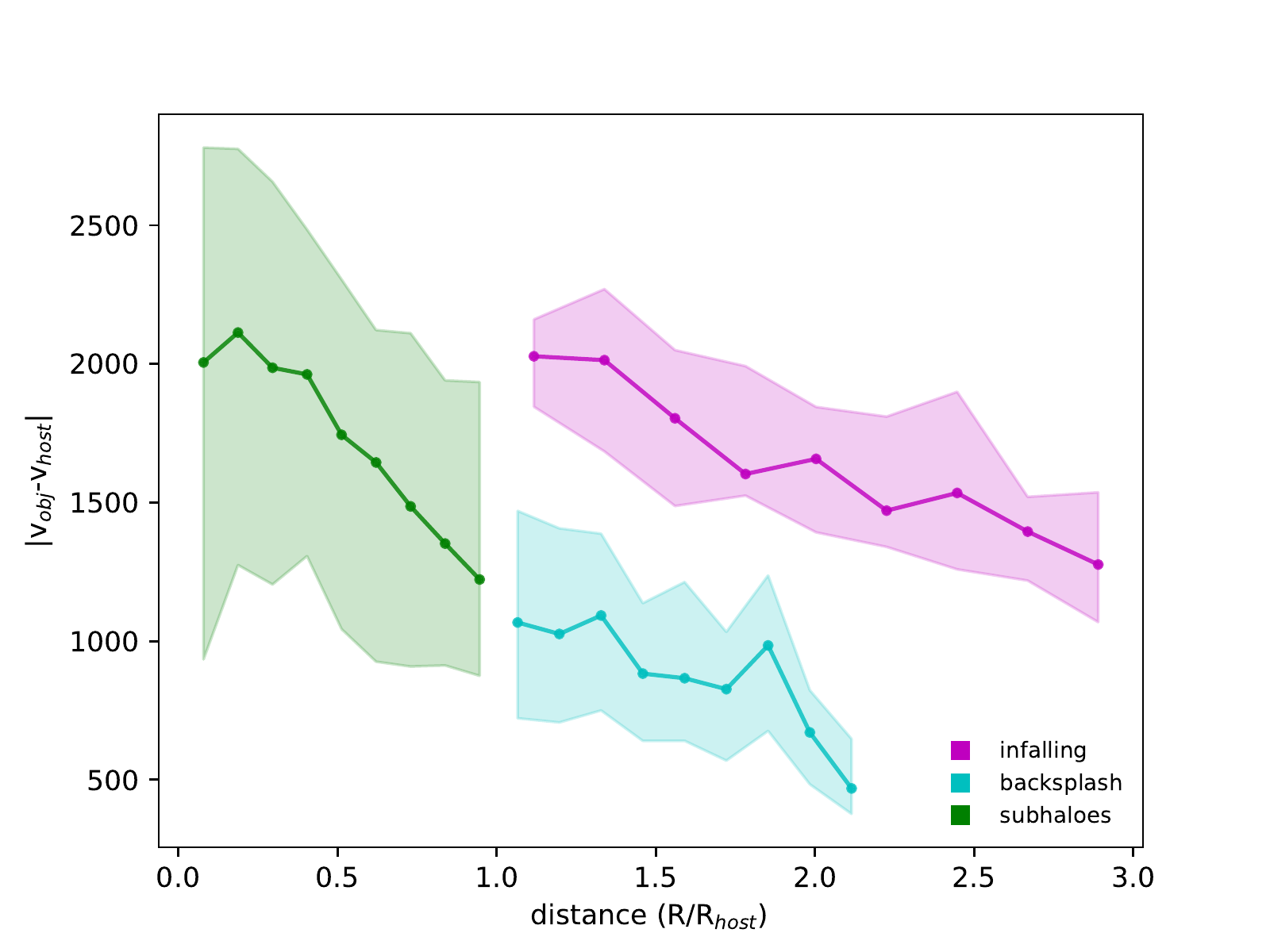}
 % \end{center}
   \caption{Halo velocity in host rest frame as a function of distance to the host.}
  \label{fig:velocity_distance}
  \end{figure}

When discussing \Fig{fig:radialalignment_distance} in \Sec{sec:RA_distance} we argued that subhaloes which reach very close to the cluster center should have large infall velocities and that they do not have time to readjust their orientation toward the center. But we also might expect an old population of objects close to the centre, having reached this far in due to dynamical friction. Additionally, their orbits should have experienced circularization (e.g. \citet{Hashimoto03}, Fig.10 in \citet{Gill04b}, \citet{Reed05b}). But in that case their (circular) velocities should be comparatively small (as can be verified by calculating the circular velocity profile for a NFW cluster-sized halo again) and they should also have their velocity vector misaligned with the position vector. The latter is excluded by \Fig{fig:velocity_position_distance} and the former by the supplementary \Fig{fig:velocity_distance} provided here. \Fig{fig:velocity_distance} shows the median velocity as a function of distance clearly indicating a substantial rise towards the centre of the host cluster.

We finally provide here a plot that compares our own data to the fitting function of \citet[][Eq. A1]{Tenneti14} when post-correcting our `spherical' shapes following \citet[][$s\rightarrow s^{\sqrt{3}}$]{Bailin05}: as can be seen, we do in fact recover the same shape-mass trend, at least for halo masses $> 10^{12.5}$\hMsun. As can be seen in the mass-distance relation provided above (\Fig{fig:MedMass_distance}), the lower mass objects are preferentially located closer to the position of the central cluster. And for those objects we have also seen that their shapes are primarily dominated by the less spherical stellar component – partly due to halo finder limitations (as seen in \Fig{fig:SMHM_distance} above). We therefore consider this deviation from the trend usually found in the literature to be artificial again. We further like to remark that the masses of the host haloes entering the Tenneti et al. work span a much larger range than ours: we are only considering subhaloes of (massive) galaxy clusters. This might have an impact on the halo masses and shapes, especially at the lower mass end: those cluster subhaloes might have experiences substantial stripping.

 \begin{figure}
 % \begin{center}
   \hspace*{-0.1cm}\includegraphics[width=8.5cm]{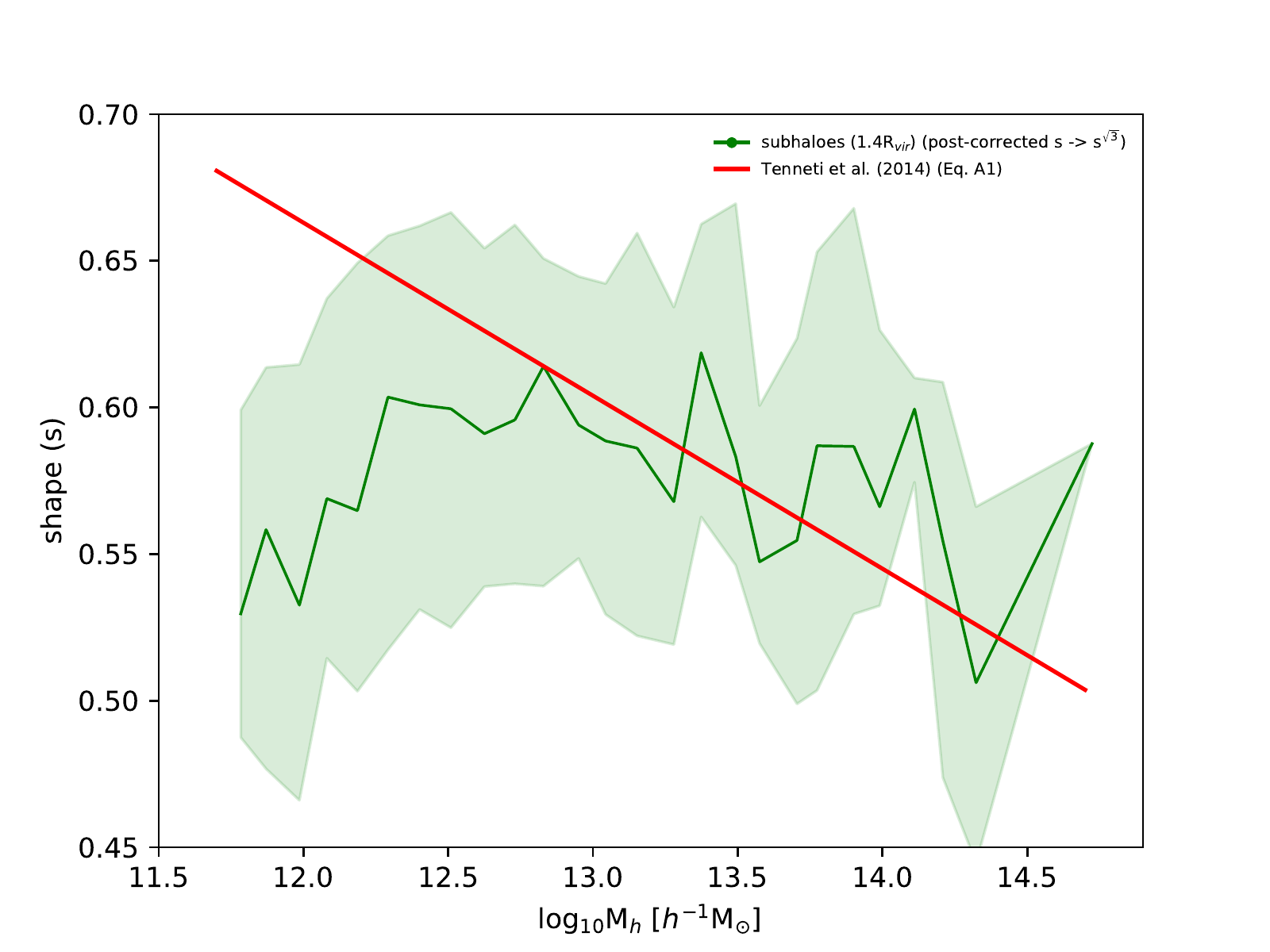}
 % \end{center}
   \caption{Post-corrected halo shape as a function of halo mass using all objects out to $1.4$\Rcrit.}
  \label{fig:ShapeMassCorrected}
  \end{figure}

%%%%%%%%%%%%%%%%%%%%%%%%%%%%%%%%%%%%%%%%%%%%%%%%%%

% Don't change these lines
\bsp	% typesetting comment
\label{lastpage}
\end{document}